\documentclass{aa}  
\usepackage{multirow}
\usepackage{natbib}
\usepackage{graphicx}
\usepackage{txfonts}
\usepackage{siunitx}
\usepackage{xcolor}
\usepackage{placeins}
\usepackage{float}
\usepackage{hyperref}
\newcommand{\jwst}{{JWST}}
\newcommand{\hst}{{HST}}
\newcommand{\um}{\SI{}{\micro\meter}}
\newcommand{\uJy}{\SI{}{\micro Jy}}
\newcommand{\orcid}[1]{\includegraphics[scale=0.06]{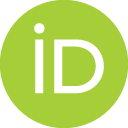} \href{https://orcid.org/#1}{#1}}
\newcommand\qth{$^{\rm th}$}

\begin{document}

     \title{MIDIS: Quantifying the AGN component of X-ray-detected galaxies}

    \titlerunning{MIDIS X-ray sources \& AGN}
    \authorrunning{S. Gillman et al.}

  \author{Steven Gillman,\inst{1,2}\thanks{\href{orcids}{ORCIDs} listed on final page}
         John P. Pye, \inst{3}
         Almudena Alonso-Herrero, \inst{4}
         Martin J. Ward, \inst{5}
         Leindert Boogaard, \inst{6}
         Tuomo V. Tikkanen, \inst{3}
         Luis Colina, \inst{7}
         G. \"Ostlin, \inst{8}
         Pablo G. P\'erez-Gonz\'alez, \inst{7}
         Luca Costantin, \inst{7}
         Edoardo Iani, \inst{9}
         Pierluigi Rinaldi, \inst{9,10}
         Javier Álvarez-M\'arquez, \inst{7} 
         A. Bik, \inst{8}
         Sarah E. I. Bosman, \inst{6,11}
         Alejandro Crespo G\'omez, \inst{7,12}
         Andreas Eckart, \inst{13}
         Macarena Garc\'ia-Mar\'in, \inst{14}
         Thomas R. Greve, \inst{1,2}
         Jens Hjorth, \inst{15}
         A. Labiano, \inst{16}
         Danial Langeroodi, \inst{15}
         J. Melinder, \inst{8}
         Florian Peißker, \inst{13}
         Fabian Walter, \inst{6}
         M. Güdel, \inst{17,18}
         Thomas Henning, \inst{6}
         P.-O. Lagage, \inst{19}
         Thomas P. Ray \inst{20}
         }
         
   \institute{Cosmic Dawn Center (DAWN), Copenhagen, Denmark\\
    \email{srigi@space.dtu.dk}
    \and
    DTU Space, Elektrovej, Building 328, 2800, Kgs. Lyngby, Denmark
    \and 
    School of Physics \& Astronomy, Space Park Leicester, University of Leicester, 92 Corporation Road, Leicester LE4 5SP, UK
    \and
    Centro de Astrobiologıa (CAB), CSIC-INTA, Camino Bajo del Castillo s/n, 28692 Villanueva de la Ca\~nada, Madrid, Spain
    \and 
    Centre for Extragalactic Astronomy, Department of Physics, Durham University, South Road, Durham DH1 3LE, UK
    \and 
    Max Planck Institut f\"ur Astronomie, K\"onigstuhl 17, D-69117 Heidelberg, Germany
    \and
    Centro de Astrobiología (CAB), CSIC-INTA, Ctra. de Ajalvir km 4, Torrejón de Ardoz, E-28850 Madrid, Spain
    \and 
    Department of Astronomy, Stockholm University, Oscar Klein Centre, AlbaNova University Centre, 106 91 Stockholm, Sweden
     \and 
    Kapteyn Astronomical Institute, University of Groningen, P.O. Box 800, 9700AV Groningen, The Netherlands
    \and 
    Steward Observatory, University of Arizona, 933 North Cherry Avenue, Tucson, AZ 85721, USA
    \and
    Institute for Theoretical Physics, Heidelberg University, Philosophenweg 12, D–69120 Heidelberg, Germany
    \and
    Space Telescope Science Institute (STScI), 3700 San Martin Drive, Baltimore, MD 21218, USA
    \and 
    Physikalisches Institut der Universität zu Köln, Zülpicher Str. 77, 50937 Köln, Germany
    \and
    European Space Agency, ESA Office, Space Telescope Science Institute, 3700 San Martin Drive, Baltimore, MD 21218, USA 
    \and
    DARK, Niels Bohr Institute, University of Copenhagen, Jagtvej 155, 2200 Copenhagen, Denmark
    \and
   Telespazio UK for the European Space Agency, ESAC, Camino Bajo del Castillo s/n, 28692 Villanueva de la Ca\~nada, Spain 
    \and
    Dept. of Astrophysics, University of Vienna, T\"urkenschanzstr. 17, A-1180 Vienna, Austria
    \and
    ETH Z\"urich, Institute for Particle Physics and Astrophysics, Wolfgang-Pauli-Str. 27, 8093 Z\"urich, Switzerland 
    \and
    AIM, CEA, CNRS, Université Paris-Saclay, Université Paris Diderot, Sorbonne Paris Cité, F-91191 Gif-sur-Yvette, France 
    \and
    Dublin Institute for Advanced Studies, 31 Fitzwilliam Place, D02 XF86, Dublin, Ireland
     }
  \abstract{X-ray and infrared surveys provide efficient, and to some degree complementary, means of detecting and characterising Active Galactic Nuclei (AGN), with the infrared also providing an important probe of the host galaxies. 
   To this end we combine the deepest X-ray survey from the Chandra Deep Field-South (CDF-S) `7-Ms' survey \citep{Luo2017} with the deepest mid-infrared (5.6~\um) image from the JWST/MIRI Deep Imaging Survey (MIDIS) in the Hubble Ultra-Deep Field (HUDF)  to study the infrared counterparts and point-source emission of 31 X-ray sources with a median, intrinsic, rest-frame X-ray luminosity of $\log_{10}$(L$_{\rm Xc}^{\rm 0.5\,-\,7\,keV})$\,=\,42.04\,$\pm$\,0.22 erg\,s$^{-1}$. The sample includes 24 AGN with a redshift range, as set by the X-ray detectability, of $z$\,$\simeq$\,0.5\,--\,3, with the bulk of the sources lying at $z$\,$\simeq$\,1\,--\,2, i.e.\ at around the epoch of Cosmic Noon. Through a multi-wavelength morphological decomposition, employing three separate classifications (visual, parametric and non-parametric) we separate (where present) the luminosity of the point-like AGN component from the remainder of the host-galaxy emission. The unprecedented mid-infrared sensitivity and imaging resolution of MIRI allows, in many cases, the direct characterisation of point-like (i.e.\ unresolved) components in the galaxies' emission. We establish a broad agreement between the three morphological classifications. At least 70\% of the X-ray sources, including some classified as galaxies, show unresolved emission in the MIRI images, with the unresolved-to-total flux fraction at rest-frame 2~\um\ ranging from $\sim 0.2$ to $\sim 0.9$. At high X-ray luminosities ($\log_{10}$(L$_{\rm Xc})$\,$>$\,43\,erg\,s$^{-1}$) we derive a consistent rest-frame near-infrared 2~\um\ point-source luminosity to that derived for local AGN, whilst at lower X-ray luminosity we identify an excess in the 2~\um\ emission compared to pre-JWST studies. We speculate this offset may be driven by a combination of Compton-thick AGN components and nuclear starburst, merger driven activity. Our observations highlight the complex nature of X-ray sources in the distant Universe and demonstrate the power of JWST/MIRI in quantifying their nuclear infrared emission.}
   \keywords{Galaxies: high-redshift, Galaxies: active, Galaxies: structure, X-rays: galaxies}

   \maketitle
%

\section{Introduction}

Since the discovery that the cosmic X-ray background was isotropic \citep{Fabian1992}, there followed major interest in determining the contributions of discrete source populations, as a function of X-ray energy. These studies combined the results from all-sky catalogues which sampled the discrete sources at limited X-ray sensitivity, with deeper serendipitous samples of sources scattered around the sky e.g.\ using  Chandra and XMM-Newton. These studies \citep{Cappelluti2017} have resolved more than 90\% of the background below 10 keV into different classes of AGN, including unabsorbed sources, partially obscured sources and Compton thick sources. At energies greater than 10 keV the fractions of the AGN population change in favour of the obscured sources.
Although there is a correlation between X-rays and the mid-infrared (MIR), it was shown that using a MIR colour selection criterion based on the WISE all sky survey \citep[e.g.][]{Stern2015}, missed a significant fraction of AGN detected in hard X-ray by NuSTAR and XMM-Newton \citep{Matoes2012}. Since the Mid Infrared Instrument (MIRI; \citealt{Wright2023}) on board JWST offers so much greater sensitivity and spatial resolution compared with all previous mid-IR facilities, it is now possible to spatially resolve high redshift ($z$\,$>$\,1) galaxies in the mid-infrared (\citealt{Boogaard2024}; \citealt{Costantin2024}).

The contribution of an AGN to a galaxy's spectral energy distribution (SED) peaks at mid-infrared wavelengths, especially for dust-obscured sources whose non-stellar emission is reprocessed through hot and warm dust emission \citep{Hickox2018}. With MIRI observations it is thus possible to isolate the unresolved AGN-heated dust emission in the rest-frame near-infrared spectral range at much higher redshifts than is possible with NIRCam \citep{Takumi2024}. At rest-frame 2\,--\,3.5~\um, type 1 (unobscured) nearby AGN show a strong correlation between the hard (2\,--\,10 keV) X-ray and the non-stellar (unresolved component) luminosities, while type 2 lie below the correlation due to torus obscuration  effects \citep[see e.g.][]{Kotilianen1992,Alonso-Herrero1997,Burtscher2015}. Thus, it is interesting to examine a sample of X-ray selected sources (AGN and galaxies, see below) detected in the MIRI Deep Imaging Survey (MIDIS; \citealt{Ostlin2024}) field. This provides new information regarding the presence or absence of an AGN dust heated nuclear component and its contribution with respect to the host galaxy emission, even in low X-ray luminosity galaxies where part of the X-ray emission may have stellar origins (e.g. X-ray binaries, shocks) and thus providing insights into the nature of the host galaxy \citep[e.g.][]{Gao2003,Barrows2024}. 

In this paper, we present a morphological analysis of a sample of 31 X-ray-selected sources in the MIDIS field with the goal of estimating the unresolved emission associated with AGN heated dust. In Section \ref{sect_xraycat} we present the parent X-ray catalogue from the Chandra 7-Ms survey before outlining the 
other observations and data reduction in Section \ref{Sec:Obs}. In Section \ref{Sec:Analysis} we define the sample of X-ray sources in the MIDIS survey and present the analysis of  multi-wavelength JWST observations with the associated results and discussion in Section~\ref{Sec:Results}, before summarising our main conclusions in Section \ref{Sec:Conc}. Throughout the paper, we assume a $\Lambda$CDM cosmology with $\Omega_{\rm m} = 0.3$, $\Omega_{\Lambda} = 0.7$, and $H_0 = 70\,\mathrm{km\,s^{-1}\,Mpc^{-1}}$. All quoted magnitudes are on the AB system and stellar masses are calculated assuming a Chabrier initial mass function (IMF) \citep{Chabrier2003}. All uncertainties reported on median values are bootstrapped uncertainties. 

\begin{figure*}
    \centering
    \includegraphics[width=\linewidth]{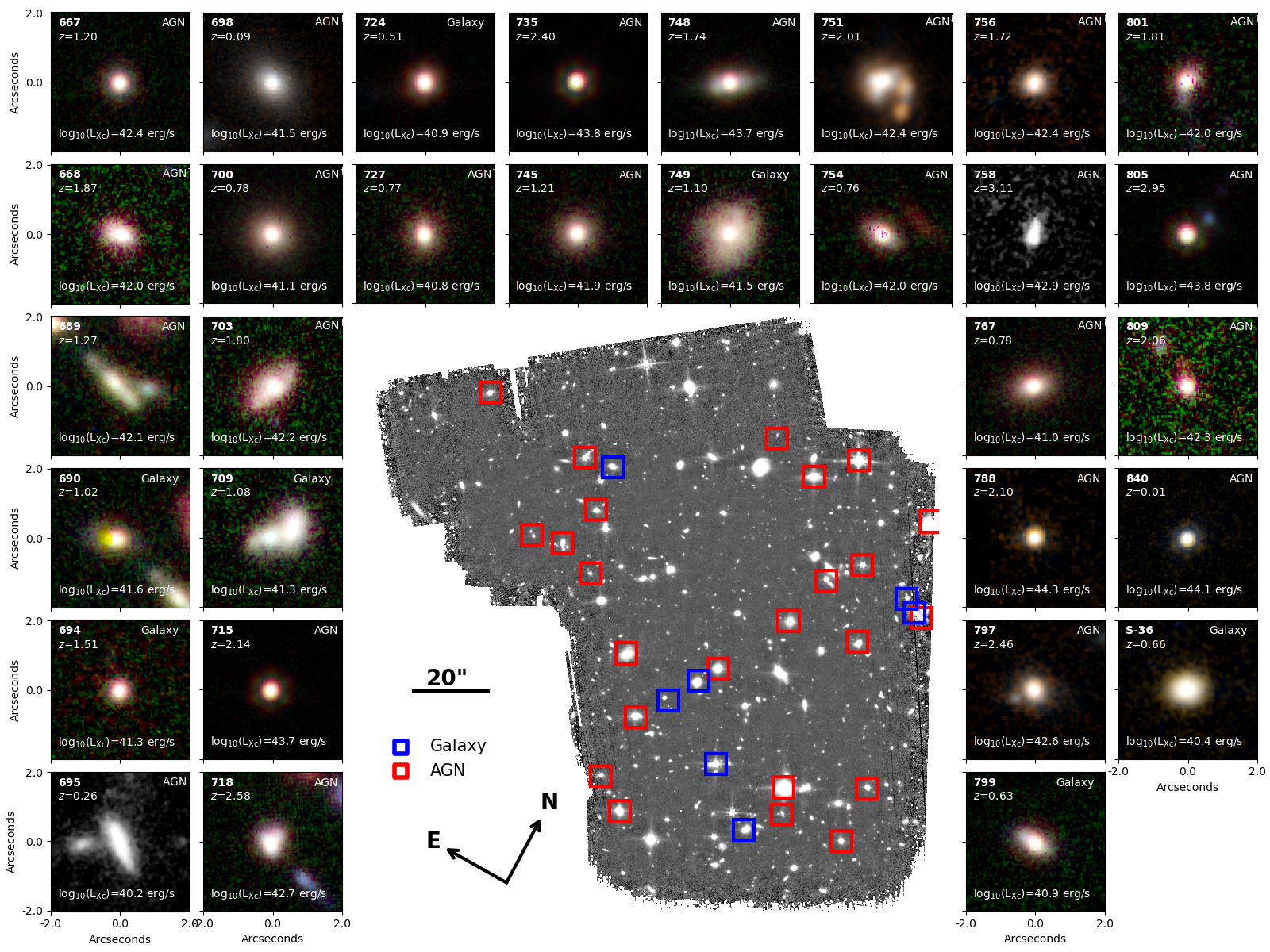}
    \caption{The MIDIS MIRI 5.6~\um\ image in the HUDF ({\it{centre}}) with 4\farcs{0}\,$\times$\,4\farcs{0} false colour maps (F1000W/F770W/F560W as R/G/B) for the 31 X-ray-detected sources. We highlight the positions of sources classified as Galaxy (blue squares) and AGN (red squares) in the 5.6~\um\ image. In each false colour image, we label the source ID from \citet{Luo2017}, spectroscopic redshift, absorption-corrected intrinsic 0.5\,--\,7.0 keV luminosity (L$_{\rm Xc)}$ and the object class (AGN or Galaxy). ID 695 and ID:758 are shown as monochromatic F560W images due to the high noise levels in the F1000W image.}
    \label{Fig:cimg}
\end{figure*}


\begin{figure}
    \centering
    \includegraphics[width=\linewidth]{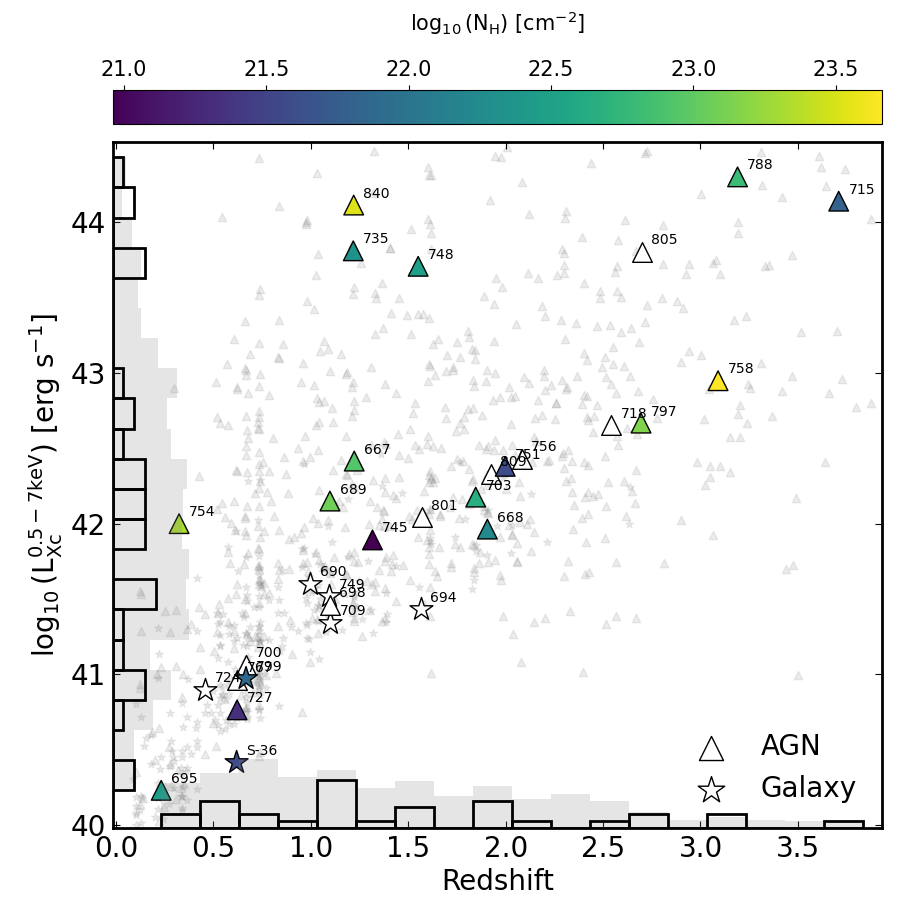}
    \caption{Absorption corrected X-ray (0.5\,--\,7 keV) luminosity as a function of redshift for the 31 X-ray sources in the MIDIS survey coloured by column density (N$_{\rm H}$). In the background we show the X-ray emitting sources identified by \citet{Luo2017} in the 7Ms survey. We show the distribution for the MIDIS sources (black histogram) and 7Ms sources (grey histogram, scaled down by a factor 10) on each axis. A clear trend of increasing X-ray luminosity with redshift is apparent, 
    as expected for a flux-limited sample.}
    \label{Fig:Lx_z}
\end{figure}


\section{The sample: CDF-S X-ray sources} \label{sect_xraycat}
We use the X-ray source catalogue from the Chandra Deep Field-South (CDF-S) `7-Ms' survey \citep{Luo2017}\footnote{Downloaded from the online CDS VizieR catalogue server: 
\url{https://vizier.cds.unistra.fr/viz-bin/VizieR-3?-source=J/ApJS/228/2} }
to identify the infrared (IR) counterparts to galaxies with detected X-ray emission in the MIDIS field. 
To date, this set of observations comprises the deepest X-ray survey conducted anywhere in the sky, and hence represents the most sensitive search for AGN X-ray emission, outside of observations utilising gravitational lensing to augment the luminosity and redshift reach (e.g.\ \citealt{Bogdan2024}). 

CDF-S covers a roughly circular region of sky, with total area 484 arcmin$^2$; it encompasses, and is roughly concentric with, the GOODS-S survey region; hence it also contains the HUDF, and thus MIDIS, which are located towards the central, most sensitive part of CDF-S. The CDF-S source survey is flux-limited, with full-band (0.5\,--\,7 keV) point-source threshold ranging from $<$\,3\,$\times$\,10$^{-17}$$\rm erg\,cm^{-2}\,s^{-1}$ close to the survey centre, to $> $\,10$^{-15}$ $\rm erg\,cm^{-2}\,s^{-1}$ near to the survey perimeter (\citealt{Luo2017}, Fig.~29). The main source catalogue contains 1008 sources above the detection-significance threshold. There is a supplementary catalogue of 47 lower-significance sources included due to proximity to bright near-infrared (NIR) candidate counterparts.
We refer to the main-catalogue sources by the running number (`XID') assigned by \citet{Luo2017}, while for the supplementary list we prefix the XID with `S-' to avoid ambiguity. As detailed in \citealt{Luo2017}, the main catalogue has a median redshift of $z$\,=\,1.2\,$\pm$\,0.05 with a 16\qth\,--\,84\qth\ percentile range of $z$\,=\,0.52\,--\,2.32. The median intrinsic (i.e.\ rest-frame), absorption-corrected, X-ray luminosity of the catalogue is $\rm \log_{10}(L_{\rm Xc}^{0.5\,-\,7 keV})$\,=\,42.5\,$\pm$\,0.06\,erg\,s$^{-1}$ with a 16\qth\,--\,84\qth\ percentile range of $\rm \log_{10}(L_{\rm Xc}^{0.5\,-\,7 keV})$\,=\,41.1\,--\,43.8\,erg\,s$^{-1}$. Whilst the median intrinsic absorption column density is $\rm \log_{10}(N_{\rm H})$\,=\,22.7\,$\pm$\,0.04\,cm$^{-2}$ with a 16\qth\,--\,84\qth\ percentile range of $\rm \log_{10}(N_{\rm H})$\,=\,21.8\,--\,23.5\,cm$^{-2}$. The Chandra images have a spatial resolution $\sim$1 arcsec, varying with photon energy and off-axis angle (e.g.\ \citealt{Primini2011}).

The source positions reported by \citet{Luo2017} are adjusted by $\Delta$RA\,=\,+128\,mas, $\Delta$Dec\,=\,$-$288\,mas, as derived by \citet{Lyu2024} for the Systematic Mid-infrared Instrument Legacy Extragalactic Survey (SMILES: PI: G. Rieke, \#1207), to bring them into alignment with the GAIA reference coordinate system adopted for the JWST (and reprocessed HST) images \citep{Gaia2021}. We cross-match the updated  Chandra positions for the X-ray sources with the footprint of the MIDIS Survey, as shown in Figure \ref{Fig:cimg}. For our analysis we require the X-ray source to be covered by at least the F560W filter in the MIDIS survey, ensuring observed-frame mid-infrared is sampled. This results in a sample of 31 X-ray sources for which \citet{Luo2017} classify 14 as `AGN', and 17 as `galaxy'. The label of `galaxy' indicates that through the multi-wavelength analysis carried out to date, no AGN component has been identified. However, using somewhat different criteria, \citet{Lyu2022}, in their (pre-JWST) AGN catalogue, classify 23 of these X-ray sources as `AGN': all those previously so classified, plus 9 from the `galaxy' category. In addition, the recent JWST/MIRI SMILES AGN survey (\citealt{Lyu2024}) yields one further AGN classification. Hence, the current tally is 24 AGN and 7 Galaxies. These AGN/Galaxy assignments are listed in Table~\ref{Table:Sample}. All 31 MIRI sources were previously detected at mid-IR (Spitzer) and near-IR/visible (HST) wavelengths; hence their basic X-ray/IR properties have already been established (e.g.\ \citealt{Luo2017}, \citealt{Lyu2022}). Six of the highest-flux  Chandra sources have X-ray spectra reported by \citet{Liu2017}. These same six sources were also detected by XMM-Newton \citep{Ranalli2013}\footnote{A seventh, low significance, source in the XMM-Newton catalogue of \citet{Ranalli2013} was noted by them as `probably spurious', and we find no reasonable MIRI-F560W counterpart.}.

For 97\% (30/31) of the MIDIS sources that originate from the main CDF-S catalogue, we compare their properties to the overall CDF-S survey as presented above. We establish a median redshift of $z$\,=\,1.3\,$\pm$\,0.2 with a 16\qth\,--\,84\qth\ percentile range of $z$\,=\,0.65\,--\,2.49. The thirty sources have a median absorption-corrected X-ray luminosity of $\rm \log_{10}(L_{\rm Xc}^{0.5\,-\,7 \,keV})$\,=\,42.1\,$\pm$\,0.21\,erg\,s$^{-1}$ and column density of $\rm \log_{10}(N_{\rm H})$\,=\,22.5\,$\pm$\,0.3\,cm$^{-2}$, with a 16\qth\,--\,84\qth\ percentile range of $\rm \log_{10}(L_{\rm Xc}^{0.5\,-\,7 \,keV})$\,=\,41.0\,--\,43.7\,erg\,s$^{-1}$ and $\rm \log_{10}(N_{\rm H})$\,=\,21.7\,--\,23.2\,cm$^{-2}$ respectively. This indicates that there is not a substantial difference in these properties between the MIDIS sample and the parent catalogue. Comparing the number of X-ray sources and the sky area for MIDIS and the full CDF-S: MIDIS occupies $\sim 1$\% of the CDF-S, but has $\sim 3$\% of the X-ray sources; this can be understood in terms of MIDIS lying in a relatively high sensitivity part of CFD-S.

\section{Observational Data}\label{Sec:Obs}

To analyse the AGN component of the X-ray detected galaxies in the MIDIS deep survey, we utilise deep multi-wavelength observations from public \hst\ and \jwst\ surveys in combination with those from the MIDIS survey of the HUDF, as we now describe.

\subsection{MIDIS} \label{Sec:midis}

A full description of the MIRI Deep Imaging Survey observations and data reduction is given in \citet{Ostlin2024}. In short, between December 2022 and December 2023, 52\,hours of F560W and 10\,hours of F1000W imaging were obtained in the HUDF. All observations were coherently reduced using a modified version of the official \jwst{} pipeline\footnote{\url{https://github.com/spacetelescope/jwst}}, version v1.12.3 (\texttt{CRDS\_CONTEXT\,=\,1137}), with additional routines to deal with cosmic-ray showers and background variations. As per the NIRCam and \hst\ observations, each exposure was aligned to Gaia DR3, co-added and drizzled to a final pixel scale of 0\farcs{04}\footnote{We note this pixel scale is different to that presented in \citet{Ostlin2024} to be consistent with the ancillary observations used in our analysis.}. In Figure \ref{Fig:cimg} we show the final 5.6~\um\ mosaic from MIDIS, with a 5$\sigma$ point-source depth of 28.6 mag (0.013~\uJy) (\citealt{Ostlin2024}), highlighting the unprecedented depth and resolution of the MIDIS observations. This is comparable to the depths of the JWST Advanced Deep Extragalactic Survey (JADES; PI: D. Eisenstein, \#1180) NIRCam observations \citep{Eisenstein2023} with a F444W 5$\sigma$ point-source depth of 29.2, whilst being significantly deeper than the wide-area observations from SMILES with a 5$\sigma$ point-source depth of 25.6 at 5.6~\um.

\subsection{HST and JWST}

We collate multi-wavelength \jwst\ and \hst\ observations in the HUDF from the DAWN JWST Archive (DJA)\footnote{\url{https://dawn-cph.github.io/dja/index.html}}, as described in Gillman et al. (in prep.). Briefly, these observations cover the full observed-frame spectral energy distribution from 0.4\,--\,4.8~\um\ with the HST (F435W, F606W, F775W, F814W) and NIRCam (F090W, F115W, F150W, F200W, F210M, F335M, F277W, F356W, F410M, F444W, F460M, F480M) filters as well as the MIRI F770W filter at 7.7~\um. The observations, taken from a variety of public \jwst\ surveys (e.g. JADES; \citealt{Eisenstein2023}, First Reionization Epoch Spectroscopically Complete Observations (FRESCO; PI: P. Oesch, \#1895) Survey; \citealt{Oesch2023}, SMILES; \citealt{Alberts2024}) were coherently reduced using the {\sc{grizli}} software \citep{Brammer2022}, which builds upon the level-2 data products from the STScI archive\footnote{\url{https://mast.stsci.edu/}} and incorporates additional routines to deal with diagonal striping, cosmic rays and stray light identified in some exposures (see \citealt{Boogaard2024, Gillman2023, Gillman2024}). We align all images to the same World Coordinate System (WCS) reference using the Hubble Legacy Field (HLF) catalogue in \citet{Whitaker2019} based on Gaia DR3 \citep{Gaia2021}. The final mosaics are co-added and drizzled to a 0\farcs{04} pixel scale \citep{Fruchter2002} for all \jwst\ and \hst\ filters.


\begin{table*}[ht!]
\centering
\caption{Properties of the 31 X-ray sources in the MIDIS MIRI field of view.}
\renewcommand{\arraystretch}{1.5}
\begin{tabular}{lccccccccc}
\hline
\hline
X-ray ID & RA & Dec. & Redshift & $\rm \log_{10}(L_{\rm Xc}^{0.5\,-\,7 keV})$ & $m_{\rm F560W}$ & $m_{\rm F770W}$ & $m_{\rm F1000W}$  & Class & MIRI Sep.\\
 & (J2000) & (J2000) & & [erg\,s$^{-1}$] & AB & AB & AB & & ["] \\
\hline
667 & 3:32:35.79 & $-$27:47:35.05 & 1.22 & 42.41 & 21.88 & 21.89 & 21.81 & AGN & 0.13 \\
668 & 3:32:35.83 & $-$27:47:19.31 & 1.91 & 41.96 & 21.86 & 22.29 & 21.97 & AGN$^\dagger$ & 0.35 \\
689$^{\rm A}$ & 3:32:36.46 & $-$27:46:31.66 & 1.10 & 42.16 & 20.82 & 20.63 & 20.75 & AGN & 0.32 \\
690 & 3:32:36.65 & $-$27:46:31.24 & 1.00 & 41.60 & 21.71 & 21.63 & 21.36 & Galaxy & 0.05 \\
694 & 3:32:36.92 & $-$27:46:29.09 & 1.56 & 41.44 & 21.99 & 22.46 & 22.62 & Galaxy & 0.29 \\
695 & 3:32:37.10 & $-$27:47:36.24 & 0.23 & 40.23 & 23.64 & 23.73 & 23.10 & AGN$^\dagger$ & 0.32 \\
698 & 3:32:37.22 & $-$27:46:07.60 & 1.10 & 41.46 & 19.45 & 19.66 & $-$ & AGN$^\dagger$ & 0.83 \\
700 & 3:32:37.34 & $-$27:47:29.49 & 0.67 & 41.08 & 19.29 & 19.86 & 20.00 & AGN$^\dagger$ & 0.37 \\
703$^{\rm A}$ & 3:32:37.39 & $-$27:46:45.42 & 1.85 & 42.18 & 21.14 & 21.53 & 21.22 & AGN$^\dagger$ & 0.53 \\
709$^{\rm A}$ & 3:32:37.65 & $-$27:47:44.61 & 1.10 & 41.34 & 21.78 & 21.67 & 22.06 & Galaxy & 0.22 \\
715$^{\rm A}$ & 3:32:38.03 & $-$27:46:26.54 & 3.71 & 44.14 & 21.31 & 20.59 & 19.84 & AGN & 0.04 \\
718$^{\rm A}$ & 3:32:38.53 & $-$27:46:34.79 & 2.54 & 42.65 & 21.66 & 21.32 & 21.40 & AGN & 0.31 \\
724$^{\rm A}$ & 3:32:38.77 & $-$27:47:32.34 & 0.46 & 40.92 & 20.27 & 19.91 & 18.19 & Galaxy & 0.16 \\
727 & 3:32:38.83 & $-$27:46:49.11 & 0.62 & 40.78 & 21.03 & 21.66 & 21.81 & AGN$^\dagger$ & 0.41 \\
735 & 3:32:39.09 & $-$27:46:02.05 & 1.22 & 43.82 & 19.37 & 19.07 & 18.89 & AGN & 0.05 \\
745 & 3:32:39.63 & $-$27:47:9.50 & 1.32 & 41.89 & 20.76 & 21.04 & 21.35 & AGN & 0.16 \\
748$^{\rm A}$ & 3:32:39.74 & $-$27:46:11.55 & 1.55 & 43.71 & 19.96 & 19.99 & 19.67 & AGN & 0.07 \\
749$^{\rm A}$ & 3:32:39.87 & $-$27:47:14.75 & 1.10 & 41.52 & 20.56 & 20.37 & 20.68 & Galaxy & 0.61 \\
751$^{\rm A}$ & 3:32:40.05 & $-$27:47:55.77 & 2.00 & 42.38 & 20.54 & 20.87 & 20.59 & AGN$^\ddagger$ & 0.32 \\
754 & 3:32:40.65 & $-$27:47:31.23 & 0.32 & 42.00 & 20.61 & 21.22 & 21.32 & AGN & 0.35 \\
756 & 3:32:40.73 & $-$27:47:49.75 & 2.08 & 42.43 & 22.17 & $-$ & 22.12 & AGN$^\dagger$ & 0.20 \\
758 & 3:32:40.75 & $-$27:46:07.16 & 3.09 & 42.95 & 24.59 & 24.57 & 24.71 & AGN & 0.93 \\
767 & 3:32:41.43 & $-$27:47:17.28 & 0.62 & 40.96 & 20.29 & 20.97 & 21.17 & AGN$^\dagger$ & 0.29 \\
788 & 3:32:42.83 & $-$27:47:02.72 & 3.19 & 44.30 & 23.01 & 36.14 & 22.24 & AGN & 0.12 \\
797$^{\rm A}$ & 3:32:43.34 & $-$27:46:46.73 & 2.69 & 42.67 & 22.25 & 22.05 & 22.20 & AGN & 0.31 \\
799 & 3:32:43.44 & $-$27:46:34.53 & 0.67 & 40.99 & 21.20 & 21.50 & 19.64 & Galaxy & 0.24 \\
801 & 3:32:43.62 & $-$27:46:58.89 & 1.57 & 42.04 & 21.74 & 21.83 & 21.60 & AGN$^\dagger$ & 0.10 \\
805$^{\rm A}$ & 3:32:44.03 & $-$27:46:35.89 & 2.70 & 43.81 & 21.27 & 20.71 & 20.22 & AGN & 0.11 \\
809 & 3:32:44.26 & $-$27:47:00.93 & 1.92 & 42.34 & 23.33 & 23.77 & 23.47 & AGN & 0.61 \\
840 & 3:32:46.34 & $-$27:46:32.26 & 1.22 & 44.11 & 22.01 & 21.38 & $-$ & AGN & 0.14 \\
S-36 & 3:32:40.23 & $-$27:47:23.21 & 0.62 & 40.41 & 23.01 & 23.47 & 21.83 & Galaxy & 1.30 \\
\hline
\hline
\label{Table:Sample}
\end{tabular}
\tablefoot{Column 1 lists the IDs adopted from \citet{Luo2017}, where an `A' indicates the source is detected in the ASPECS ALMA survey of the HUDF \citep{Boogaard2024}. Columns 2 and 3 indicate the X-ray-source coordinates, adjusted as described in Sect.~\ref{sect_xraycat}. The updated spectroscopic redshifts of the sources are listed in column 4. The absorption-corrected intrinsic 0.5\,--\,7.0 keV luminosity, amended where necessary (from that reported by \citealt{Luo2017}) to reflect updated redshift, is given in column 5. The MIDIS 5.6~\um, SMILES 7.7~\um\ and MIDIS 10~\um\ integrated magnitudes are given in columns 6, 7 and 8. The class (AGN or Galaxy) is given in column 9 as defined in \citet{Luo2017} and F560W to X-ray offset is given in column 10. Objects classified as Galaxy in \citet{Luo2017} but as AGN in \citet{Lyu2022} or \citet{Lyu2024} are indicated with a $\dagger$ or $\ddagger$ respectively.}
\end{table*}


\begin{figure*}
    \centering
    \includegraphics[width=\linewidth,trim={0cm, 0.7cm, 0cm, 0cm},clip]{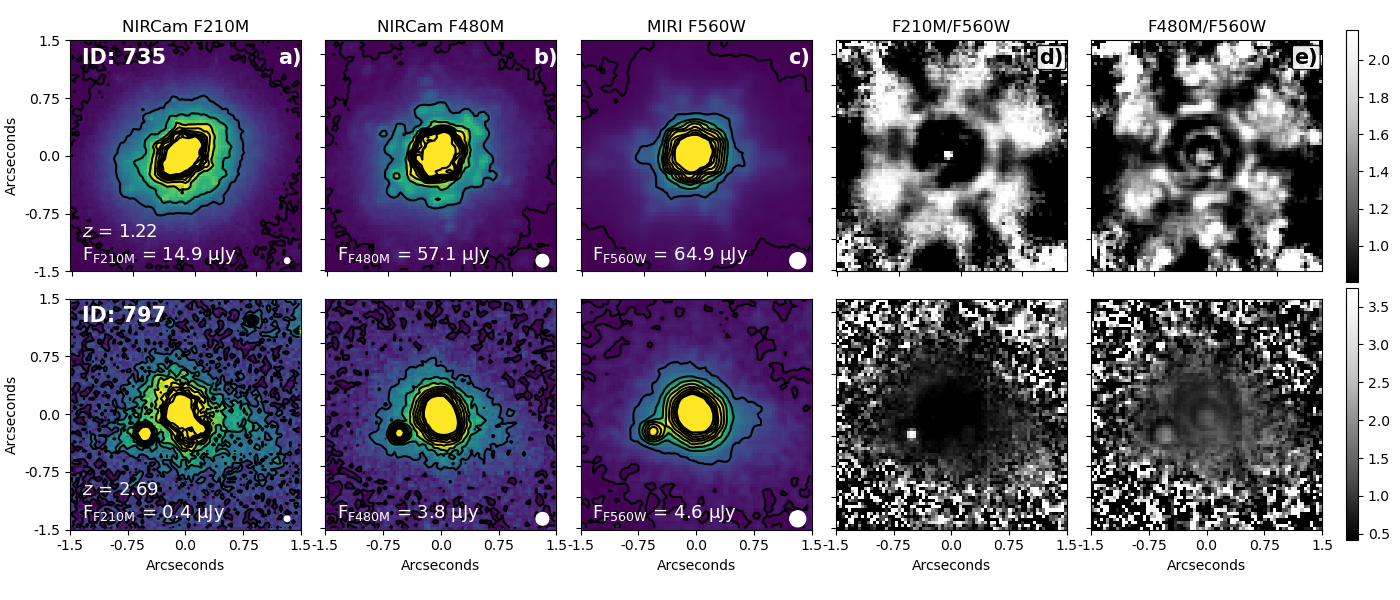}
    \caption{JWST multi-wavelength imaging and colour maps for two examples of the X-ray sources in our sample. The full sample is shown in Appendix \ref{App:CMaps}. For each galaxy, we show, and contour, a 3-arcsecond cutout in the NIRCam F210M band (a), the NIRCam F480M band (b) and the MIRI F560W band (c). We label the galaxy's ID as given by \citet{Luo2017}, measured flux in each band and spectroscopic redshift. We also indicate the FWHM of the PSF in each band by the white circle in the lower-right corner. The F210M/F560W (d) and F480M/F560W (e) colour maps are shown for each source, highlighting the presence of any AGN (point source) component at longer (near-infrared) wavelengths.}
    \label{Fig:colmaps}
\end{figure*}

\section{Analysis}\label{Sec:Analysis}

To study the properties of the point-source (AGN) component of the X-ray detected sources in the MIDIS survey, we first identify the MIRI F560W and F1000W counterparts and characterise their multi-wavelength properties prior to employing a two-dimensional spatial decomposition routine to separate the AGN component and host galaxy.

\subsection{MIRI Counterparts}\label{subsec:MIRI}

To identify the MIRI counterparts, we cross-match the sample of 31 X-ray sources, with updated positions (see Section \ref{sect_xraycat}) to {\sc{The Farmer}} catalogue, as presented in Gillman et al. (in prep.). {\sc{The Farmer}} catalogue, which utilises the deep F560W image from the MIDIS survey as the detection image, employs a model-based forced photometry approach to define the multi-wavelength fluxes of the detected sources from 0.4\,--\,10~\um. 

To identify a MIRI counterpart to each of the 31 sources in the MIDIS field of view, we require a counterpart search radius of 1\farcs{5}. As shown in Figure \ref{Fig:cimg}, we identify X-ray sources across the full extent of the MIDIS footprint, with a median X-ray to 5.6~\um\ offset of $\delta$r\,=\,0\farcs{25}\,$\pm$\,0\farcs{05} arcseconds and a 16\,--\,84\qth\ percentile range of $\delta$r\,=\,0\farcs{11}\,--\,0\farcs{48}. 
Comparing $\delta$r for each source with the corresponding X-ray positional error radius (1 standard deviation ($\approx 68$\% confidence level), $\sigma_{\rm r}$, \citealt{Luo2017}), we find that all the sources lie within $\delta$r$/\sigma_{\rm r}$\,$<$\,2.0 (i.e.\ within the 95\% confidence level), with only three, rather weak, sources (ID:698, ID:758 and ID:S-36) exceeding $\delta$r$/\sigma_{\rm r}$\,$\approx$\,1.2. These same three sources also have relatively large ($\sim 1$ arcsec) counterpart-to-X-ray offsets in \citet{Luo2017}. All 31 X-ray sources have clearly identified MIR counterparts, with a median {\sc{Farmer}} model-based 5.6~\um\ magnitude of $m_{\rm F560W}$\,=\,21.3\,$\pm$\,0.3 with a 16\qth\,--\,84\qth\ percentile range of $m_{\rm F560W}$\,=\,20.3\,--\,22.4 and median 5.6~\um\ S/N\,=\,1199\,$\pm$\,264. The full properties of the sample are presented in Table \ref{Table:Sample}. In addition to the optical and infrared detections from HST and JWST respectively, 11 sources (35\%) of the sample are detected in the far-infrared as part of the ALMA Spectroscopic Survey in the Hubble Ultra Deep Field (ASPECS; PI: F. Walter, \citealt{Walter2016}) as highlighted in Table \ref{Table:Sample}. For a complete analysis of the ALMA-detected sources in the MIDIS survey see \citet{Boogaard2024}.

\subsection{Redshifts}

There are spectroscopic redshifts for all 31 sources in our sample.
For ninety per cent of them (28/31), spectroscopic redshifts are available from the far-infrared surveys \citep[e.g.][]{Boogaard2020,Boogaard2024} as well as near-infrared optical studies such as the MUSE HUDF Survey \citep[][]{Bacon2023}. For the remaining three sources without published spectroscopic redshifts we utilise the public spectroscopic observations from FRESCO \citep{Oesch2023} that target the MIDIS field. Following the same procedure as \citet{Boogaard2024}, we extract spectroscopic redshifts for ID:797 at $z$\,=\,2.695 and ID:805 at $z$\,=\,2.698 based off the detected nebula line emission (e.g. H$\alpha$, [OIII]). For the remaining source, ID:695, which is not covered in the FRESCO Survey, we search the Next Generation Deep Extragalactic Exploratory Public (NGDEEP, PI: S. Finkelstein, \#2079) Survey \citep{Bagley2024}. Again adopting the procedure of \citet{Boogaard2024}, we identify a spectroscopic redshift of  $z$\,=\,1.56 for the X-ray source ID:694.

Our sample has a median redshift of $z$\,=\,1.22\,$\pm$\,0.19 with a 16\qth\,--\,84\qth\ percentile range $z$\,=\,0.62\,--\,2.48. The redshifts of the full sample are presented in Table \ref{Table:Sample}. For 12 sources we identify a different spectroscopic redshift to the photometric redshift used by \citet{Luo2017}. We therefore update the redshift, and intrinsic X-ray luminosity, for the sources in our sample, with a median ratio of $\rm L_{Xc}^{MIDIS}$/L$_{\rm Xc}^{\rm L17}$\,=\,1.04\,$\pm$\,0.05. The updated X-ray luminosities, as well as spectroscopic redshifts, for the full sample are reported in Table \ref{Table:Sample} and in Figure \ref{Fig:Lx_z} we plot the 0.5\,--\,7 keV absorption-corrected X-ray luminosity (L$\rm_{Xc}$) for the 31 X-ray sources in the MIDIS survey as a function of their redshift, showing the comparison to the full sample presented in \citet{Luo2017}.

\subsection{Morphological Analysis}

In Figure \ref{Fig:cimg} we show the MIRI F560W/F770W/F1000W false colour images for the X-ray sources, highlighting the diverse range of morphology in the sample from compact, point-source red objects to extended, disc-like systems. These MIRI mid-infrared observations sample the galaxy emission at rest-frame $>$1~\um\, providing insights into the structure and morphology of the continuum emission. 

We visually assess the morphologies of the X-ray sources, in addition to both a non-parametric and parametric analysis of the MIRI F560W and F1000W observations, to quantify their rest-frame near-infrared morphology. We exclude the \hst\ and NIRCam observations from our morphological analysis because we are interested in identifying the AGN component of the galaxies, which becomes more prominent at wavelengths greater than rest-frame 1~\um. The analysis of the \jwst\ imaging employs 5\farcs{0}\,$\times$\,5\farcs{0} cutouts of each source with pixel scales of 0\farcs{0}4 per pixel\footnote{We use a slightly larger cutout than shown in Figure \ref{Fig:cimg} and Figure \ref{Fig:colmaps} to ensure the background is well modelled.}. The cutout in each filter is centred on the source detected in the MIDIS 5.6~\um\ observation. 
We use {\sc{photutils}} \citep[][]{Phot2022} to model (and remove) the background level in each cutout as well as to quantify the root-mean-square (rms) noise. In the following sections "cutout" refers to these  10\farcs{0}\,$\times$\,10\farcs{0}, background subtracted images that are used in the morphological analysis that follows.

Before measuring the galaxies' morphology, we first define the point spread function (PSF) for each of the \jwst\ bands. We adopt the PSFs used in Gillman et al. (in prep.) which were used to build {\sc{The Farmer}} catalogue. For the F560W band, this includes an empirical PSF made from stacked stars that accurately models the extended cross-form pattern of the F560W PSF, whilst for the MIRI F1000W band the PSF is generated from  {\sc{WebbPSF}} \citep{Perrin2014}. For full details see Gillman et al. (in prep.).

\begin{figure*}
    \centering
    \includegraphics[width=\linewidth]{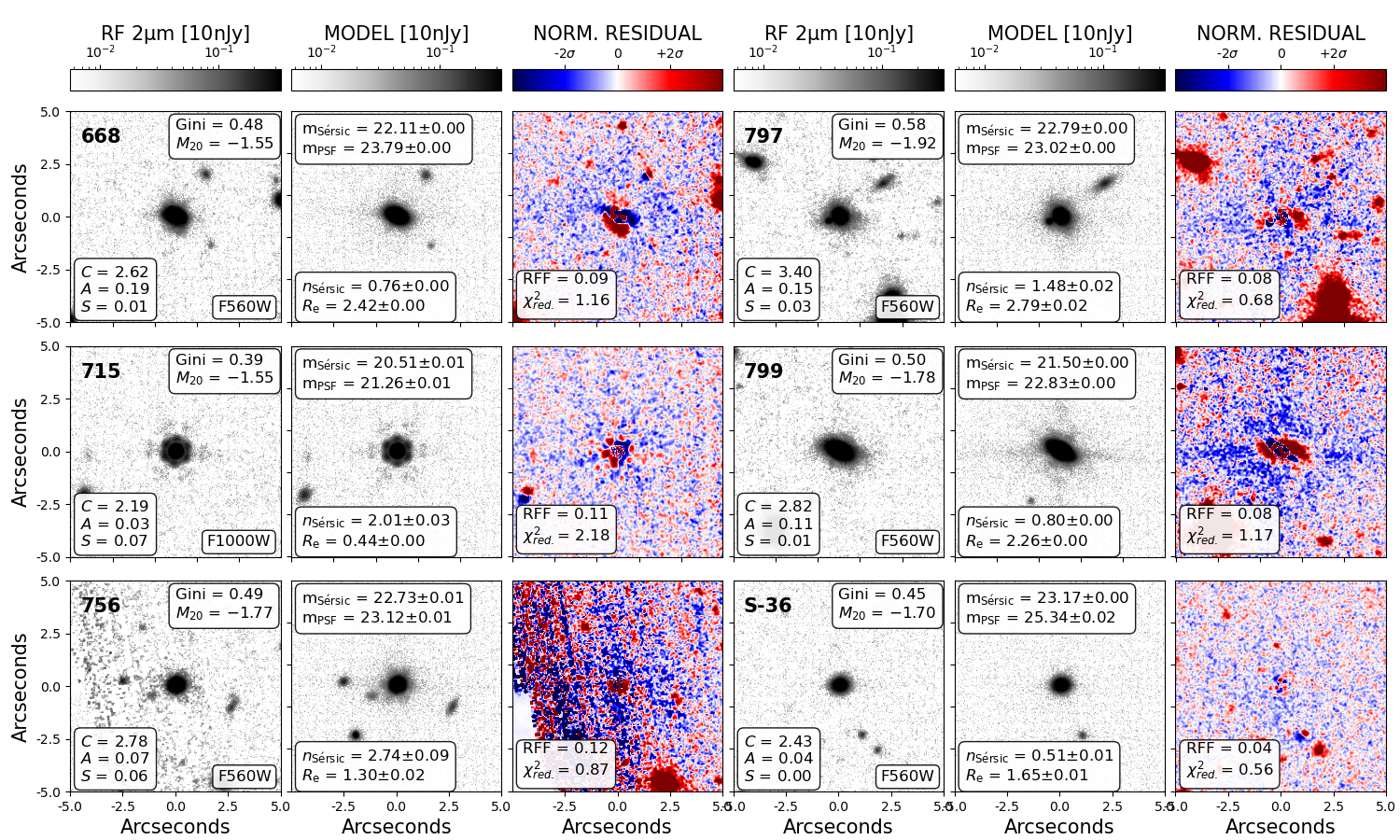}
    \caption{Examples of the rest-frame 2~\um\ morphological analysis for six X-ray sources in our sample, highlighting the diverse range of morphology in the sample. For each source, we show a 10-arcsecond cutout, best-fit  S\'ersic plus PSF parametric model and  normalised residuals (residual/$\sigma$). We label the source ID, MIRI filter, non-parametric (Gini, $M_{20}$, $C$, $A$, $S$)  and parametric ($n_{\rm S\grave{e}rsic}$, $R_{\rm e}$) morphological parameters, as well as goodness of fit parameters ($\chi^2_{\rm red.}$, RFF). For each parametric model we also annotate the S\'ersic and PSF derived magnitude.}
    \label{fig:morph_example}
\end{figure*}

\subsubsection{Visual Morphology} \label{Sec:Vis}
To aid the visual classification we also generate two colour maps for each galaxy, an example of which is shown in Figure \ref{Fig:colmaps}. The colour maps contrast the NIRCam F210M and F480M filters, which sample the rest-frame optical emission of the galaxies, with the MIRI F560W observation, which samples the rest-frame near-infrared emission. In this way, by contrasting the rest-frame optical and near-infrared emission any unresolved AGN-heated dust emission that begins to dominate over the stellar emission in the near-infrared clearly stands out \cite[e.g.][]{Martini2003}. The colour maps of the full sample are shown in Appendix \ref{App:CMaps}. 

The visual classification of all 31 target galaxies utilising the MIRI F560W and F1000W ``cutouts'' as well as corresponding colour maps was carried out by six team members (and co-authors). The initial classifications were assigned independently, and a final, merged outcome was produced from an interactive discussion. Where deemed necessary, a more detailed inspection was done with the F560W and F1000W images displayed using DS9\footnote{\url{https://sites.google.com/cfa.harvard.edu/saoimageds9/home}} and interactively optimising the greyscale or colour scale to view the faintest or brightest regions of an object.

For ease of comparison with similar studies, we adopt the classification scheme used by \citet{Kartaltepe2023} in studying galaxy structure and morphology at $z$\,=\,3\,--\,9 from the Cosmic Evolution Early Release Science (CEERS: PI: S.Finkelstein, \#1345) programme. Hence we allow for recognition of up to four categories of surface-brightness morphology: {\it Disk}, {\it Spheroid}, {\it Irregular}, and {\it Point-source/Unresolved}. The first three are collectively referred to as {\it Extended}. Each galaxy can exhibit multiple extended-morphology classes.  The main aim of this exercise is to identify the presence of: (i) point-source components, and (ii) extended emission components, for comparison with the morphological results from the parametric and non-parametric methods. We caution that there is some degree of subjectivity and uncertainty concerning the visual classification into the three individual extended-emission categories, as discussed in \citet{Kartaltepe2023}, nevertheless, we consider that they form a useful basis for the cross-comparisons.

\begin{figure*}
    \centering
    \includegraphics[width=\linewidth]{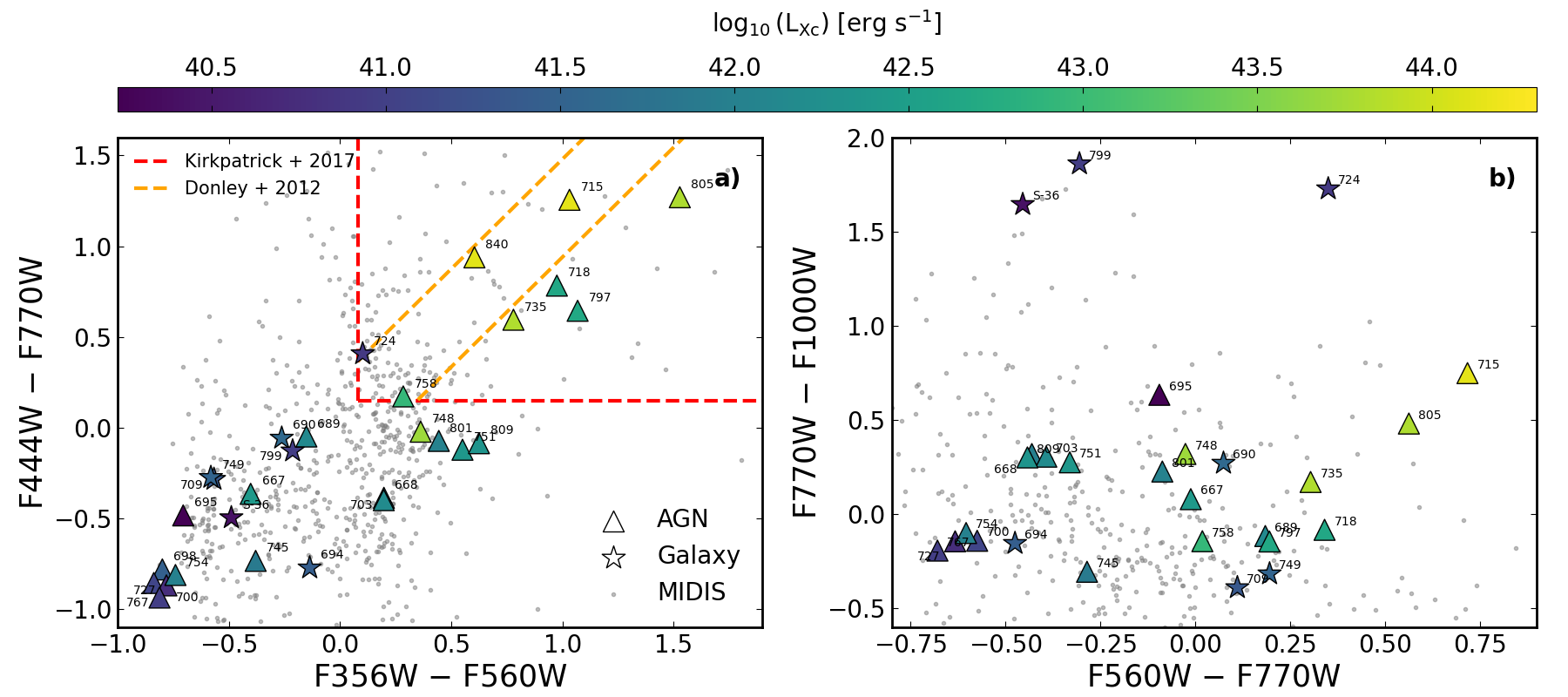}
    \caption{The near-infrared colour properties of the 31 X-ray sources coloured by X-ray luminosity. Grey points are star-forming galaxies in the MIDIS field of view above $z$\,$>$\,0.5. We show the F444W\,$-$\,F770W colour versus F356W\,$-$\,F560W colour $(a)$, highlighting the near-infrared colour selected AGN region proposed by \citet{Kirkpatrick2017} (red-dashed line) and \citet{Donley2012} (orange-dashed line). In panel $b)$ we plot the MIRI colours, F560W\,$-$\,F770W versus F770W\,$-$\,F1000W.  The three X-ray emitting galaxies with F770W\,$-$\,F1000W\,$>$\,1.5, have a colour likely driven by the presence of the 6.2~\um\ PAH feature in the F1000W filter with redshifts between $z$\,=\,0.46\,--\,0.67 as demonstrated by \citet{langeroodi2023}.}
\label{Fig:MIRI_colours}
\end{figure*}


\subsubsection{Non-Parametric Modelling}

Next, we adopt a non-parametric approach to quantify the morphology of the X-ray-selected galaxies. This approach invokes no assumptions on the underlying structure of the galaxy's light distribution, thus highlighting the unique properties of each source. Prior to quantifying the morphology we first smooth the source segmentation map generated from {\sc{The Farmer}} using the binary dilation routine in {\sc{photutils}}. Then, excluding this segmented region, we mask the remaining sources in the cutout, down to a 1$\sigma$ isophote. This ensures full masking of any contaminants (spurious or otherwise). We run {\sc{statmorph}}\footnote{https://statmorph.readthedocs.io/en/latest/} \citep{statmorph2019} on the masked cutouts in the 5.6~\um\ and 10~\um\ bands using the same segmentation maps and PSFs as for the {\sc{the farmer}} catalogue (see Section \ref{subsec:MIRI}). 

{\sc{statmorph}} measures the Concentration, Asymmetry and Clumpiness ($C$, $A$, $S$; \citealt{Abraham2003,Lotz2008}) parameters that quantify how concentrated, asymmetrical and clumpy the galaxies' surface brightness profiles are, with higher values indicating more concentrated, more asymmetric, or clumpier light profiles respectively.  
In addition, the  Gini and M$_{20}$ parameters are also derived (for full definitions see \citealt{Lotz2004,Synder2015}). Briefly, the Gini parameter defines the pixel distribution of the galaxy's light, where $G$\,=\,1 corresponds to all of the light  concentrated in one pixel whilst $G$\,=\,0 indicates each pixel contributes equally. The M$_{20}$ parameter measures the second moment of the brightest 20 percent of pixels in the galaxy. This is normalised by the total  moment for all pixels.  Highly negative values indicate a high concentration of light, not necessarily at the centre of the galaxy. 

\subsubsection{Parametric Modelling}

Finally, to quantify both the parametric morphological profiles of the galaxies and the deviations from these, we use the \texttt{\sc{galfitm}} code \citep{Haubler2013}. \texttt{\sc{galfitm}} is a Python-wrapper for \texttt{\sc{galfit}} \citep{Peng2010}, that allows multi-component parametric models to be fit to a galaxy's multi-wavelength light distribution. For our analysis we fit each band independently thus allowing us to constrain the intrinsic wavelength dependence of each galaxy's morphology. We further simultaneously model, with a single S\'ersic model \citep{Sersic1963}, all sources within 3\farcs{0} of the X-ray source, as identified in the {\sc{The Farmer}} segmentation map (see Section \ref{subsec:MIRI}) .

To model the target source, on which the cutout is centred, we fit two different models to each band to determine the presence of a point-source component to each galaxy's light profile. First, we fit a single free S\'ersic model with magnitude, S\'ersic index ($n$), half-light radius, position angle and axis ratio free to vary. We then fit a free S\'ersic model plus PSF model with the PSF model defined from the input PSF and parametrised with a position angle and magnitude. An example of the parametric modelling of the galaxies is shown in Figure \ref{fig:morph_example}.

To evaluate the goodness of fit of each of the models, we employ two metrics. First, we quantify the significance of the residuals to each model in each band using the residual flux fraction (RFF), as defined in \citet{Hoyos2011,Hoyos2012},
\begin{equation}
    \rm RFF\,=\,\frac{\sum_{i,j \in A}|I_{i,j}\,-\,I_{i,j}^{model}|\,-\,0.8\,\times\,\sum_{i,j \in A}\,\sigma_{Bkg\,i,j}}{\sum_{i,j \in A} I_{i,j}}
\end{equation}
where the sum is performed over all pixels within 2.5 times the Kron radius as derived in the F560W image. $\rm |I_{i,j}\,-\,I_{i,j}^{model}|$ is the absolute value of pixel $i,j$'s residuals to the model S\'ersic fit, whilst $\sum_{i,j \in A}I_{i,j}$ indicates the total flux measured in the source as defined in Section \ref{subsec:MIRI}. The 0.8 factor, multiplied by the sum over the background rms of the region ($\sigma_{Bkg\,i,j}$), ensures that a blank image with a constant variance has an RFF\,=\,0.0 (see \citealt{Hoyos2012} for details).  Finally, we calculate the reduced chi-squared of each model which is defined as:
\begin{align}
    \chi^{2}\,&= \, \sum_{i}^{N_{\rm d}} r_{i}^{2} \\
    \chi^{2}_{\rm red.}\,&= \, \chi^{2} / (N_{\rm  d}\,- \,N_{\rm  varys})
\end{align}
where $\displaystyle\sum^{N_{\rm d}}_{i}r_{i}^{2}$ is the sum of the residual image, $N_{\rm d}$ is the number of data points and $N_{\rm  varys}$ is the number of variable parameters. The model with the $\chi^{2}_{\rm red.}$ close to unity is preferred.

\subsection{Aperture photometry of point-source components}

As an independent verification of the point-source fluxes derived from the {\sc{galfitm}} parametric modelling, which assumes the point-source component is well modelled by the PSF model, we perform aperture photometry on the target galaxies in the F560W image and their free single S\'ersic model residual images. We expect the {\sc{galfitm}} modelled point source fluxes to be larger than those derived from the residuals whilst being smaller than those derived from the input images that contain contributions from the host galaxy.

We use {\textsc{sep}} \citep{Barbary2016}, a pythonic version of \textsc{source extractor} \citep{Bertin1996} to perform aperture photometry in 0\farcs{2} radius circular apertures centred on the MIRI source. We correct the aperture fluxes to ``total" following the aperture corrections as reported by STScI\footnote{\url{https://jwst-docs.stsci.edu/jwst-mid-infrared-instrument}}. For the residuals, on average, we find $\log_{10}(\rm L_{PSF}^{GALFIT})/\log_{10}(\rm L_{PSF}^{Aperture})$\,=\,1.04\,$\pm$\,0.01. On the images themselves we derive a median value of $\log_{10}(\rm L_{PSF}^{GALFIT})/\log_{10}(\rm L_{PSF}^{Aperture})$\,=\,0.98\,$\pm$\,0.02 highlighting the contributions from the host galaxy to the point-source emission.

As a final validation of the point-source fluxes, we re-run the {\sc{galfit}} modelling using the MIRI ePSF models from \citet{Libralato2024}. We derive a field of view varying ePSF model following the same procedure as for the empirical PSF (see \citet{Boogaard2024} for full details). For the 31 X-ray sources we derive a lower point-source flux when using the ePSF models ($f_{\rm ePSF}/f_{\rm empircal}$\,=\,0.70\,$\pm$\,0.03), which we suspect is driven by the lower noise level in the ePSF model and brighter cruciform pattern at large radii when compared to the empirical PSF. We note however the choice of PSF does not alter the results presented in the forthcoming sections.

\section{Results and Discussion} \label{Sec:Results}

From our sample of 31 X-ray detected galaxies, we have 24 AGN/AGN-candidates and 7 star-forming galaxies as classified from previous literature studies \cite[e.g,][]{Luo2017,Lyu2022,Lyu2024} with a median redshift of $z$\,=\,1.22\,$\pm$\,0.19 and a 16\qth\,--\,84\qth\ percentile range $z$\,=\,0.62\,--\,2.48. The AGN sources are identified through a multi-wavelength analysis that encompasses the objects' X-ray, radio and infrared properties. One such selection is the mid-infrared colour selection, which isolates the potential AGN contribution to the near-infrared emission \citep[e.g.][]{Lyu2024}. 

In Figure \ref{Fig:MIRI_colours}a, we show the source integrated F444W\,$-$\,F770W versus F356W\,$-$\,F560W colour-colour diagram which mimics the IRAC colour selection commonly used to isolate AGN sources \citep{Donley2012,Kirkpatrick2017}, as indicated by the red and orange-dashed region. Whilst a small fraction (7/31) of the X-ray detected sources fall in the AGN parameter space, isolating the most luminous X-ray sources in our sample, the majority do not. It is well known that this colour selection is not robust against obscured or high-redshift AGN that can have near-infrared colours that mimic star-forming galaxies due to the dominance of stellar emission at these wavelengths in these systems \citep[e.g.][]{Cardamone2008,Donley2012,Mateos2015,Kirkpatrick2017,Lyu2022}.

In Figure \ref{Fig:MIRI_colours}b, we show the MIRI F560W, F770W, F1000W colour--colour diagram, highlighting the comparable brightness across the three filters. In three of the X-ray sources classified as star-forming galaxies we identify excess F1000W emission compared to F770W with an F770W\,$-$\,F1000W\,$>\,$1.5. These three galaxies (ID:724, ID:799 and ID:S-36) have redshifts between $z$\,=\,0.46\,--\,0.67, making the F1000W filter sensitive to the 6.2~\um\ feature indicating the presence of star-formation activity. We suspect this drives their excess F1000W emission, as highlighted in \citet{langeroodi2023}. This excess is not seen in the AGN at similar redshifts due to the presence of non-stellar emission dominating over the PAH features.

\begin{table}
\centering   
\caption{Visual classification statistics of the 31 X-ray emitting galaxies detected in the MIDIS survey.}             
\label{Table:VisStats}      
\begin{tabular}{lll|l }     
\hline      
Frequency                & Likely &  &  \\
                         & Present & Present & Interacting \\
\hline                    
Disc                     & 6    & 7       &                      \\
Spheroid                 & 11    & 12       &                \\
Irregular                & 8    &   2       &                  \\
\hline                  
Extended Sub-total       & 10 & 17 & 14 \\
Point Source (Diff. pattern)             & 8  & 14 (10) & \\
\hline
\end{tabular}
\tablefoot{Multiple extended-component categories may be present in a single object.}
\end{table}


\subsection{Visual Morphology}

A summary of the visual classifications is given in Table~\ref{Table:VisStats}, with a detailed breakdown of each source given in  Appendix \ref{sect_visClass}. For a point-source component, we identify eight (6 AGN / 2 galaxy) objects where we consider it is likely present, 14 (11/3) where it is definitely present, and of these latter, ten (9/1) where the characteristic pattern of the instrumental PSF \citep{Perrin2014} can be seen. 
For the nine cases where we do not report the presence of a point source, this should be taken as `absence of evidence' rather than `evidence of absence', as complexity and contrast in a galaxy's emission morphology can limit the visual classification process. For extended emission, there are ten (8/2) objects where we consider it is likely present, and 17 (13/4) where it is definitely visible. There are 17 (12/5) objects with possible multiple extended component types, and 14 (11/3) where the morphology indicates a possible interacting system.

\subsection{Morphological Modelling}

To evaluate the presence of a point-source component by modelling the surface brightness profiles of the X-ray emitting galaxies, we assess the relative goodness of fit of parametric models using the reduced chi-squared and RFF parameters. For each galaxy we calculate a $\Delta \chi^2$\,=\,$\chi^2_{\rm S\grave{e}rsic}$\,$-$\,$\chi^2_{\rm S\grave{e}rsic+PSF}$ and $\Delta$RFF\,=\,RFF$_{\rm Single}$\,/\,RFF$_{\rm Single+PSF}$. Defined in this way, a galaxy with $\Delta \chi^2$\,$>$\,0 and $\Delta$RFF\,$>$\,1 indicates its morphology is better modelled with a single S\'ersic plus PSF model.

On average, for the 31 galaxies in the F560W filter we identify an improved reduced chi-squared when fitting a S\'ersic plus PSF model with a median value of $\chi^2_{\rm S\grave{e}rsic}$\,$-$\,$\chi^2_{\rm S\grave{e}rsic+PSF}$\,=\,3.35\,$\pm$\,1.12 (see Appendix \ref{App:GALFIT}). We further establish an average reduction in the amplitude of residuals for the two component model with $\Delta$RFF\,=\,1.28\,$\pm$\,0.08. In two sources ID:\,695 and ID:\,709, we identify an increase in the residuals when fitting two components. Upon visual inspection (see Appendix \ref{sect_visClass}) the morphological measurements of both galaxies are clearly impacted by a companion, possibly merging, galaxy.

\begin{figure}
   \includegraphics[width=\linewidth]{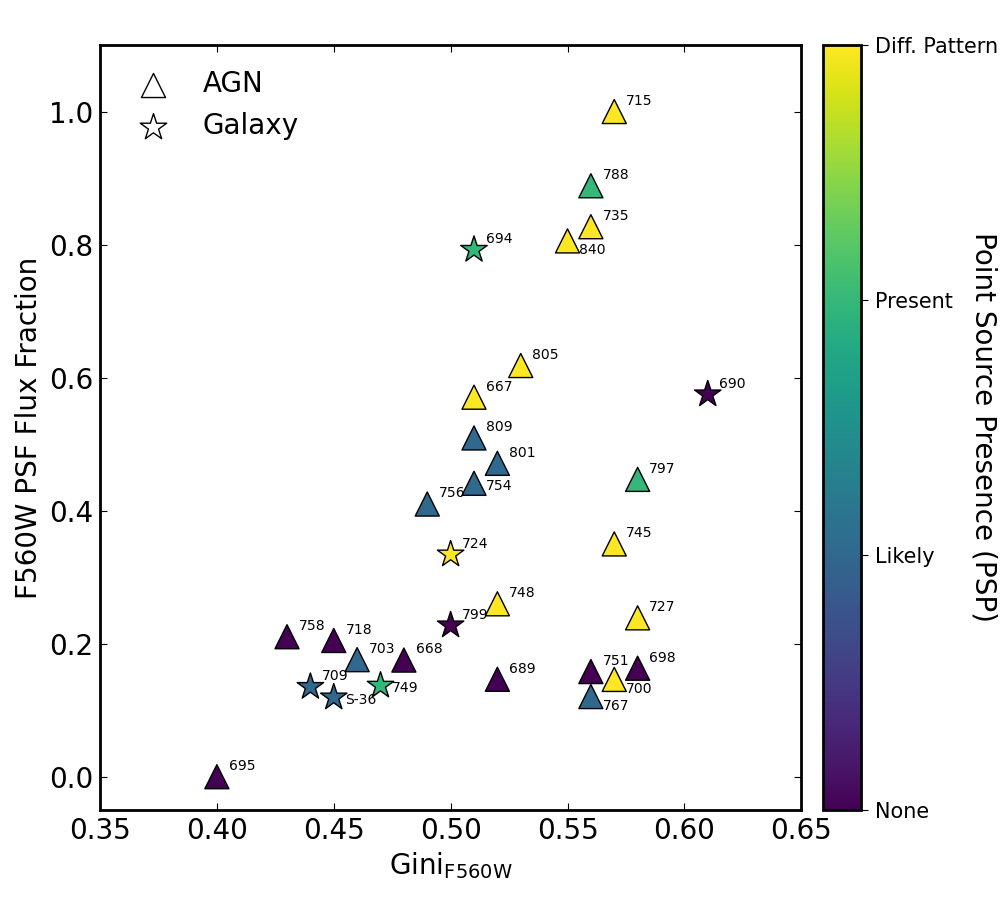}
    \caption{Example comparison of the three morphological analysis methods, for the MIRI/F560W waveband, for each source:
    Gini ({\sc{statmorph}}) vs.\ fractional point-source flux density $\rm ff_{PSF}$ (\texttt{\sc{galfitm}}), with the visual classifications of point-source presence indicated by colour-coding of each symbol ({\it purple} -- no point-source identified; {\it blue} -- likely; {\it green} -- yes; {\it yellow} -- yes, with a visible PSF characteristic diffraction pattern).
    The source classification, `AGN' or `Galaxy', is indicated by symbol shape: 
    {\it triangle} and {\it star} respectively.
    The numbers by each symbol give the X-ray source reference (XID, \citealt{Luo2017}).
    }
\label{Fig:F560W_Gini_fracFluxPsf}
\end{figure}


In Figure~\ref{Fig:F560W_Gini_fracFluxPsf} we show the fractional point-source flux density in the F560W filter, $\rm ff_{PSF}^{F560W}$, as a function of the Gini parameter (Gini$^{\rm F560W}$), where $\rm ff_{PSF}$ is the point-source flux density relative to the total (S\'ersic + PSF) flux density in the {\sc{galfitm}} model. The sources are coloured by the visual classification of point source presence (PSP) as defined in Sect.~\ref{Sec:Vis}).We note there is overall, a positive correlation (as might be expected) between Gini and $\rm ff_{PSF}$, albeit with a broad dispersion, indicating good agreement between the different morphological parameters. Those `AGN' with visually no point source presence have a $\rm ff_{PSF} \la 0.2$, and similarly for the sources classified as `Galaxy'  with one exception ($\rm ff_{PSF} \sim 0.55$). Conversely, it appears that point-source components with $\rm ff_{PSF} \ga 0.25$ can be reliably discerned by \texttt{\sc{galfitm}}. All sources with Gini$\ga 0.5$ have the PSF pattern visible. 

A recent study by \citet{Bonaventura2024}, which analysed the near-infrared morphology of a sample of mid-infrared selected AGN, identified a 3-$\sigma$ correlation between the observed-frame 1.5~\um\ asymmetry and the column density of the sources, establishing that mergers are more common amongst high-redshift obscured AGN. For the 24 AGN in our MIDIS sample of X-ray selected sources, we find no strong correlation between the 5.6~\um\ asymmetry and column density, with a median asymmetry of $A_{5.6~\um}$\,=\,0.06\,$\pm$\,0.02, below the `mild' disturbance threshold of $A$\,=\,0.1 \citep[see][]{Bonaventura2024}. We note that whilst we find no correlation between asymmetry and column density in our sample, this is not a direct comparison given the difference in rest-frame wavelength probed between the two studies. Furthermore, our sample is X-ray selected whilst the \citet{Bonaventura2024} AGN sample includes (X-ray undetected) infrared selected AGN which from previous HST studies, have been shown to exhibit a higher fraction of mergers \citep[e.g.][]{Kocevski2015,Donley2018,Ji2022}.

\begin{figure*}
    \centering
    \includegraphics[width=\linewidth]{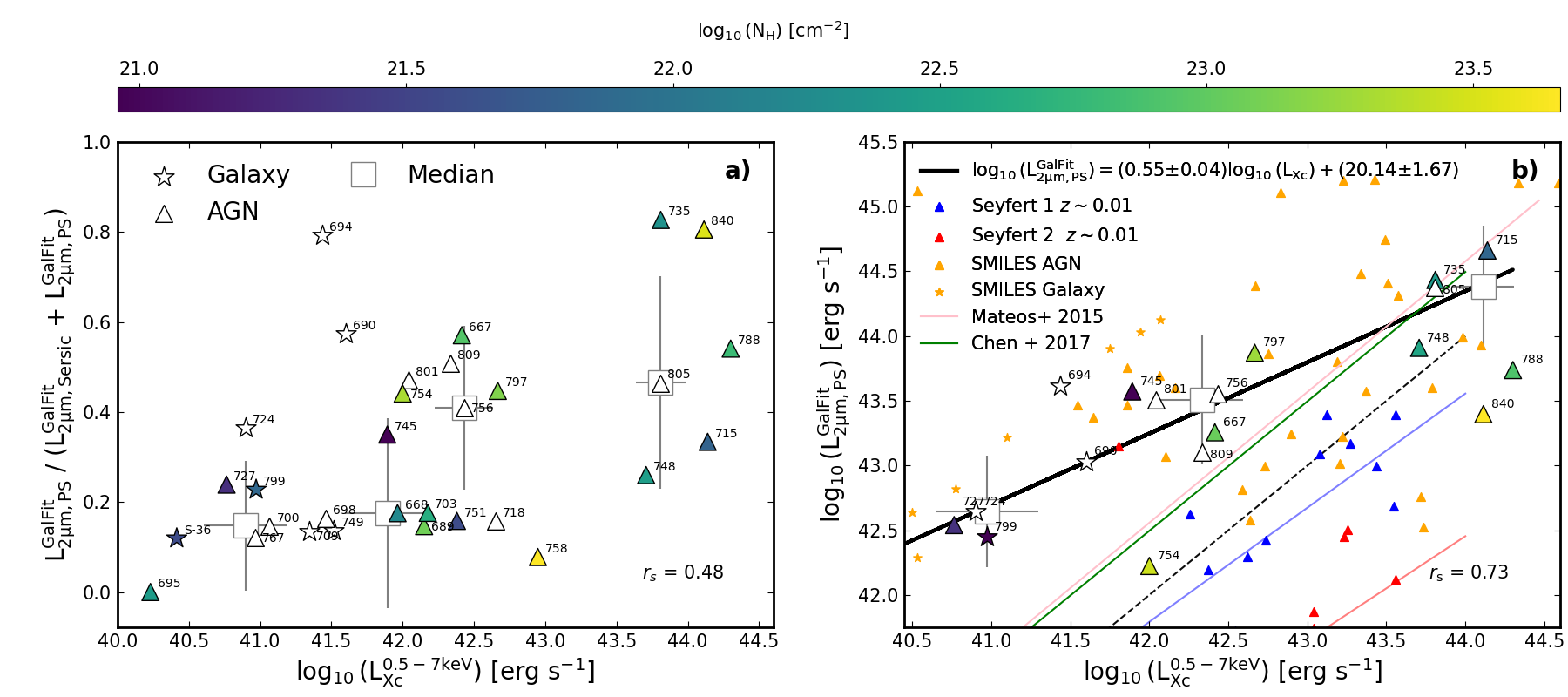}
    \caption{We show the relation between the ratio of rest-frame 2~\um\, point source (PS) luminosity to total (PSF + S\'ersic) 2~\um\ luminosity {\it{(a)}} and PS luminosity {\it{(b)}} as a function of the 0.5\,--\,7.0 keV X-ray luminosity. In each panel we show the Spearman rank correlation coefficient ($r_{\rm s}$) for the relations, a running median (grey squares) in bins of 1\,dex in X-ray luminosity. We exclude sources in panel b) with a PS to total luminosity ratio of less than 20\%.  We further show the one-to-one relation (black-dashed line) and overlay nearby ($z$\,$\sim$\,$0.01$) AGN from \citet{Burtscher2015}, as well as the 6~\um\,--\,X-ray luminosity relations from \citet{Mateos2015} for $z$\,$\simeq$\,$0.2\,$--$\,2.8$ AGN and higher redshift Quasars from \citet{Chen2017}, as well as the MIRI-SMILES X-ray sources from \citet{Lyu2024}.
    Whilst we identify a strong ($r$\,$=$\,0.73) correlation between near-infrared point-source luminosity and X-ray luminosity in our samples, at $\log_{10}$(L$_{\rm Xc})$\,$\leq$\,43 erg\,s$^{-1}$, we identify higher 2~\um\ PS luminosities than other studies, with no strong correlation with column density (N$_{\rm H}$).}
\label{Fig: L_MIRv_Lx}
\end{figure*}


\subsection{Comparison with X-ray luminosity}
Given the broad range of redshift of our sample from  $z$\,=\,0.2\,--\,3.6, the rest-frame wavelength probed by the MIDIS 5.6~\um\ (10~\um) observations varies from 4.7~\um\ (8.3~\um) to 1.2~\um\ (2.1~\um). To align the observations to a common rest-frame wavelength, we select the observation that samples closest to the rest-frame 2~\um\ emission. At 2~\um\, the emission is mostly unaffected by recent star-formation and highlights any AGN component if present. Given the redshift range detailed in Table \ref{Table:Sample}, for 84\% (26/31) of the sources we select the F560W filter, for 6\% (2/31) we use the F1000W filter and for the remaining 10\% (3/31) we use the F770W observations from the MIRI-SMILES survey as detailed in Section \ref{Sec:Obs}. In doing so, on average we sample the rest-frame 2.54\,$\pm$\,0.16~\um, emission with a 16\qth\,--\,84\qth\ percentile range of 1.94\,--\,3.47~\um. 

To convert the point-source luminosity of the galaxies, as derived from {\sc{galfitm}}, to an intrinsic rest-frame luminosity we first convert the observed-frame fluxes to luminosities using the luminosity distance of each source. We then apply a $K$-correction of the form (1+$z$) following the prescriptions of \citet{Hogg1999} to derive the rest-frame 2~\um\ luminosity. Finally we scale the luminosities by the rest-frame frequency of the observations to give final units of erg\,s$^{-1}$. In Figure \ref{Fig: L_MIRv_Lx}a, we correlate the relative contribution of the point source component to the total 2~\um\ luminosity of each source, with the absorption-corrected intrinsic 0.5\,--\,7.0 keV X-ray luminosity, identifying a correlation with a Spearman rank coefficient of $r_{\rm s}$\,=\,0.48. For 13/31 (42\%) of the sources we find the relative point-source contribution to the total luminosity to be less than 20\%. These sources exhibit no correlation with X-ray luminosity and represent the fainter X-ray sources in our sample with $\log_{10}$(L$_{\rm Xc})$\,$\leq$\,43 erg\,s$^{-1}$. For sources with a point-source contribution above 20\%, we identify a stronger correlation with X-ray luminosity, but no dependence on the column density (N$_{\rm H}$), which we adopt from \citet{Luo2017}.

Of the galaxies in our sample, 5/7 (71\%) have a point source to total luminosity ratio below 40\%, with one galaxy (ID:\,695) with the lowest X-ray luminosity in our sample ($\log_{10}$(L$_{\rm Xc})$\,=\,40.23 erg\,s$^{-1}$) having almost zero contribution from a point source in the mid-infrared ($<$0.01\%), indicating the X-ray emission may arise from X-ray binaries or merger driven activity in the extended (non-nuclear) regions of the galaxy \citep[e.g.][]{Gao2003,Barrows2024}. In contrast, two X-ray sources classified as galaxies (ID:\,694 and ID:\,690) have very high levels of point-source contribution, 79\% and 57\% respectively. Both of these galaxies are also visually classified as possibly undergoing interactions or mergers which makes the morphological modelling more uncertain. Furthermore both galaxies have small half-light radii ($\rm R_h$\,$<$\,0\farcs{26}), comparable to the PSF size of the MIRI observations, making it challenging to resolve the presence of a host galaxy, with both sources remaining point-source like even at rest-frame optical wavelengths ($\lambda$\,$<$\,0.4~\um). As noted by \citet{Luo2017}, some of the X-ray sources classified as galaxies in their analysis might host low-luminosity or heavily obscured AGN that remain undetected based on the six selection criteria used for AGN identification (see \citet{Luo2017} for full details.). Looking at the spectral properties of ID:\,694 and ID:\,690 from the NGDEEP prism spectra (see Section \ref{Sec:Analysis}) there is no clear evidence of AGN activity i.e broad-line emission. 

In Figure \ref{Fig: L_MIRv_Lx}b we correlate the 2~\um\ point-source luminosity of the sources with their X-ray luminosity. We exclude sources with point-source to total fraction of less than 20\%  (13 sources) which showed no correlation with X-ray luminosity as identified in Figure \ref{Fig: L_MIRv_Lx}a.  We identify a strong correlation with a Spearman rank coefficient of $r_{\rm s}$\,=\,0.73, although no strong dependence on the column density as derived by \citet{Luo2017}. We derive a running median in bins of 1\,dex in X-ray luminosity, as shown by the grey squares in Figure \ref{Fig: L_MIRv_Lx}b, which we quantify further using a linear orthogonal distance regression (ODR) fit of the form,
\begin{equation}
     \log_{10}(\rm L_{2~\um,PS}^{GalFit})\,=\,\alpha\log_{10}(\rm L_{Xc})+\beta
\end{equation}
establishing $\alpha$\,=\,0.55\,$\pm$\,0.04 and $\beta$\,=\,20.10\,$\pm$\,1.70.

We compare our relation between near-infrared point-source luminosity and X-ray luminosity to other literature samples of Quasars and AGN. Specifically we plot the sample of local ($z$\,$\sim$\,0.01) Seyfert 1 (blue triangles) and 2 (red triangles) AGN from \citet{Burtscher2015}. We convert the 2\,--\,10 keV X-ray luminosities reported in \citet{Burtscher2015}, to the 0.5\,--\,7 keV energy band using a scaling factor of 0.73 derived from 
{\sc{webPIMMS}}\footnote{\url{https://heasarc.gsfc.nasa.gov/cgi-bin/Tools/w3pimms/w3pimms.pl}},
assuming a (negative) power-law spectrum with photon index of 1.8 and (galactic) column density absorption of $\rm N_H$\,=\,$9 \times 10^{19}$\,$\rm cm^{-2}$ following \citealt{Luo2017}.\footnote{For an index of 1.4, the conversion factor would be $\approx$\,1.0.} We note the near-infrared luminosity reported in \citet{Burtscher2015} is derived from a combination of the radial CO equivalent-width profile and spectral decomposition, and might thus not be directly comparable to our morphological decomposition.

In addition to the local AGN, we compare to the relations derived by \citet{Mateos2015} and \citet{Chen2017}, for X-ray AGN and Quasars across a broad range of redshift ($z$\,$\simeq$\,0.2\,--\,2.8). Both these studies report the relation between the galaxy integrated rest-frame mid-infrared (6~\um) and 2\,--\,10 keV X-ray luminosity, which we convert to the 0.5\,--\,7 keV energy band using the same scaling as for the \citet{Burtscher2015} sample. We further include the comparison to the recent study by \citet{Lyu2024} of the 217 AGN reported in the MIRI SMILES survey in GOODS-S. In Figure \ref{Fig: L_MIRv_Lx}b we plot the X-ray detected MIRI-SMILES AGN  (orange triangles) and Galaxies (orange stars), again converting to the 0.5\,--\,7 keV energy band. All three of these samples are derived at longer rest-frame wavelengths than 2~\um.
In the absence of strong star-formation activity, at 6~\um\ for a power-law spectrum any AGN component of the X-ray source will dominate over the stellar emission of the host galaxy, removing the need for any decomposition of fluxes at this wavelength. Thus we expect the rest-frame 6~\um\, luminosity to be larger than the 2~\um\, luminosity in our sample.

In Figure \ref{Fig: L_MIRv_Lx}b we see that, whilst the high X-ray luminosity ($\rm \log_{10}(L_X)$\,$\geq$\,43.5 erg\,s$^{-1}$) MIDIS sources are consistent with other studies, the lower luminosity sources are offset to higher near-infrared luminosities, compared to all pre-JWST surveys. However the low X-ray luminosity MIDIS sources are comparable to the 6~\um\,--\,X-ray luminosity relation identified in the MIRI-SMILES survey, with near-infrared luminosities more than an order of magnitude higher than expected from the \citet{Burtscher2015} sample. Whilst the low X-ray luminosity MIDIS sources align with the MIRI-SMILES galaxies and AGN from \citet{Lyu2024}, as noted before this is not a direct comparison to the SMILES sample due to the offset in rest-frame wavelength probed by the infrared observations.

To understand the origin of this offset we investigate a number of different scenarios for the X-ray sources classified as galaxies and AGN. Firstly, the origin of the near-infrared point-source emission might come from nuclear starburst activity. For instance, known local nuclear starbursts such as Arp 220, with an X-ray luminosity of $\approx$\,$10^{40}$\,erg\,s$^{-1}$ has a rest-frame 2.2~\um\ point-source luminosity of  $\approx$\,$10^{43}$\,erg\,s$^{-1}$ \citep[e.g.][]{Scoville2000,Iwasawa2011}. Of the three most offset sources in this regime, two (ID:\,799 and ID:\,724) are classified as galaxies, indicating the AGN contribution is minimal or currently undetected. As also highlighted in Figure \ref{Fig:MIRI_colours}, these two X-ray sources have redshifts of $z$\,$\sim$\,0.6, thus placing the 3.3~\um\ PAH feature in the F560W filter. Hence this may result in an excess in this filter thus boosting the near-infrared luminosity in the presence of strong star-formation activity \citep[e.g.][]{langeroodi2023}. Furthermore, ID:\,724 has a reported 1-mm flux of $S_{\rm 1.2 mm}$\,=\,34\,$\pm$\,10\,\uJy\, from the ASPECS survey \citep{Aravena2020, Boogaard2024}, further indicating the presence of intense star-formation activity. Both ID:799 and the AGN with the largest offset (ID:727) have featureless NGDEEP prism spectra (see Section \ref{Sec:Analysis}), with no clear signs of AGN activity. Meanwhile ID:724, has a clear detection of a broad Pa$\beta$ line, potentially indicating the presence of AGN activity in this source previously classified as a galaxy. We note however it is difficult to definitively determine the presence of AGN activity in the prism spectrum given the low ($\sim$2000 km\,s$^{-1}$) resolution.

Another possible reason for the large offset at low X-ray luminosities, might be that the intrinsic X-ray luminosity of these sources could be underestimated if their column density is underestimated. We can calculate the increase in column density required for the least luminous X-ray sources, with $\rm \log_{10}(L_{\rm Xc})$\,$\sim$\,41.0 erg\,s$^{-1}$, to be consistent with the \citet{Burtscher2015} sample. For example, the AGN source ID:\,727 has a power-law index of $\gamma$\,=\,1.65 and a column density of $\rm N_H$\,=\,2.1\,$\times$\,10$^{21}$\,cm$^{-2}$ as reported by \citet{Luo2017}.
Using {\sc{webPIMMS}}, we estimate that a column density of $\rm N_H$\,$\sim$\,3\,$\times$\,10$^{24}$\,cm$^{-2}$ would be needed to yield the required absorption factor $\sim$\,30 to be applied to the observed X-ray luminosity. We note that this scenario may not be physically reasonable, given that the source has a significant detection in the Chandra `soft' band (0.5\,--\,2 keV; \citealt{Luo2017}). However, in the next luminosity bin at  $\rm \log_{10}(L_{\rm Xc})$\,$\sim$\,42.04 erg\,s$^{-1}$ the required column density to enhance the intrinsic X-ray luminosity is not so extreme, with the uncertainty on the median bin making it nearly consistent with the relations from \citet{Mateos2015} and \citet{Chen2017}.

We note that the excess identified at low X-ray luminosity is unlikely to be driven by just one of these scenarios, and is more likely a relative combination of all processes that is driven by the unique evolutionary path of each galaxy. This study of the X-ray sources within the MIDIS observations has demonstrated the power of deep, high resolution mid-infrared imaging with MIRI whilst highlighting the complex nature of X-ray sources in the distant Universe and the need for the upcoming MIDIS-Red observations at 7.7~\um\ and 10~\um\ with equivalent 5$\sigma$ point source depth, that will unveil the true nature of X-ray sources and AGN in the HUDF.


\section{Conclusions}\label{Sec:Conc}

The unprecedented mid-IR sensitivity and imaging resolution of MIRI, and the great depth of the MIDIS survey, allow, in many cases, the direct characterisation of point-like (i.e.\ unresolved) components in high-redshift galaxy images, at rest-frame near-infrared wavelengths.

\begin{itemize}
    \item Our sample consists of the 31 X-ray sources (CDF-S, \citealt{Luo2017}) in the MIRI Deep Imaging Survey (MIDIS) of the HUDF, using two filters F560W and F1000W.  
    \item From the measured mid-infrared emission, the sources classified as `Galaxy' do not appear to have clearly distinct properties, though we note that three of them show high colour values F770W\,$-$\,F1000W, likely driven by the presence of the 6.2~\um\ PAH feature in the F1000W filter with redshifts between $z$\,=\,0.46\,--\,0.67 (Fig.~\ref{Fig:MIRI_colours}b).
    \item We employ three methods to examine the presence of rest-frame mid-infrared unresolved  emission, which is assumed to arise in AGN-heated hot dust. Two are based on morphological modelling techniques (parametric and non-parametric), and one on visual inspection. There is broad agreement between these methods (Fig. \ref{Fig:F560W_Gini_fracFluxPsf}). 
    \item At least 70\% of the sources show unresolved emission in the MIRI images, with the unresolved-to-total flux fraction ranging from $\sim$\,0.2 to $\sim$\,0.9. 
    About half of our sample have published X-ray column densities \citep{Luo2017}. We find no strong correlation between the column densities and the ratio of the point-source component to the total 2-micron rest-frame luminosity. For an obscured AGN having a high column density, we might have expected a smaller point-source contribution. 
    \item In common with previous studies we confirm a strong correlation between the 2-micron (rest-frame) point-source luminosity and the X-ray luminosity, with a Spearman rank coefficient of $r_{\rm s}$\,=\,0.73 (Fig. \ref{Fig: L_MIRv_Lx}). We establish a shallower slope than previous studies with $\alpha$\,=\,0.55\,$\pm$\,0.04, giving rise to an excess rest-frame near-infrared 2~\um\ luminosity below $\rm \log_{10}(L_{\rm Xc})$\,=\,43.5 erg\,s$^{-1}$, comparable to that derived by \citet{Lyu2024} at 6~\um. 
    \item We speculate this offset to be driven by a combination of compact galaxy size, nuclear starburst activity, Compton-thick AGN and 3.3~\um\ PAH emission acting to enhance the near-infrared point source emission.
\end{itemize}

The current MIDIS data will be augmented during 2024--2025 by deep imaging at longer wavelengths (F770W and F1000W) with programme id.\ 6511, PI G.~Östlin,
facilitating identification of obscured AGN at redshifts $\sim $3--4.


\begin{acknowledgements}
The observations analysed in this work are made with the NASA/ESA/CSA James Webb Space Telescope (DOI: \hyperlink{https://archive.stsci.edu/doi/resolve/resolve.html?doi=10.17909/z7p0-8481}{10.17909/z7p0-8481}). SG acknowledges financial support from the Villum Young Investigator grants 37440 and 13160 and the Cosmic Dawn Center (DAWN), funded by the Danish National Research Foundation (DNRF) under grant No. 140.
JPP and TVT acknowledge financial support from the UK Science and Technology Facilities Council, and the UK Space Agency.
JH and DL were supported by research grants (VIL16599, VIL54489) from VILLUM FONDEN.
AE and FP acknowledge support through the German Space Agency DLR
50OS1501 and DLR 50OS2001 from 2015 to 2023.
AAH acknowledges support from grant PID2021-124665NB-I00  funded by MCIN/AEI/10.13039/501100011033 and by ERDF `A way of making Europe'.
ACG acknowledges support by grant PIB2021-127718NB-100 from the Spanish Ministry of Science and Innovation/State Agency of Research MCIN/AEI/10.13039/501100011033, and by JWST contract B0215/JWST-GO-02926.
JAM acknowledges support by grant PIB2021-127718NB-100 funded by MCIN/AEI/10.13039/501100011033 and by ERDF `A way of making Europe'. 
G\"O, AB and JM acknowledge support from the Swedish National Space Administration (SNSA).
SEIB is supported by the Deutsche Forschungsgemeinschaft (DFG) under Emmy Noether grant number BO 5771/1-1.
PGP-G acknowledges support from grant PID2022-139567NB-I00 funded by Spanish Ministerio de Ciencia, Innovaci\'on y Universidades MICIU/AEI/10.13039/501100011033, and the European Union FEDER program {\it Una manera de hacer Europa}.
Cloud-based data processing and file storage for this work are provided by the AWS Cloud Credits for Research program. The data products presented herein were retrieved from the Dawn JWST Archive (DJA). DJA is an initiative of the Cosmic Dawn Center, which is funded by the Danish National Research Foundation under grant No. 140.
This research has made use of the SIMBAD database, the VizieR catalogue access tool, and the Aladin sky atlas, developed by and operated at CDS, Strasbourg, France.
This research has made use of the Astrophysics Data System (ADS), funded by NASA under Cooperative Agreement 80NSSC21M00561.
For the purpose of open access, the authors have applied a Creative Commons Attribution (CC BY) licence to the Author Accepted Manuscript version arising from this submission.

\end{acknowledgements}

\section*{Facilities}
ALMA, Chandra, HST, JWST, XMM-Newton \\

\section*{Software}
DS9 (\url{https://ds9.si.edu/doc/user/index.html}), 
jwst science calibration pipeline (\url{https://jwst-docs.stsci.edu/jwst-science-calibration-pipeline#gsc.tab=0}), 
cds [aladin, simbad, vizier] (\url{https://cds.unistra.fr/}),
ads (\url{https://ui.adsabs.harvard.edu/}),
Astropy \citep{Astropy2013},  
Photutils \citep{Phot2022}, 
Source Extractor \citep{SE1996},
SEP \citep{Barbary2016},
Eazy-py \citep{Brammer2021},
GriZli \citep{Brammer2022},
GalfitM \citep{Haubler2013},
Statmorph \citep{statmorph2019},
Topcat \citep{Topcat2005},
STILTS \citep{Stilts2006},
{\sc{the farmer}} \citep{Weaver2023},
WebbPSF \citep{Perrin2014} \\

\hypertarget{orcids}{\section*{ORCIDs}}

Steven Gillman \orcid{0000-0001-9885-4589}\\
John P. Pye \orcid{0000-0002-0932-4330}\\
Almudena Alonso-Herrero \orcid{0000-0001-6794-2519}\\
Martin J. Ward \orcid{0000-0003-1810-0889}\\
Leindert Boogaard \orcid{0000-0002-3952-8588}\\
Tuomo V. Tikkanen \orcid{0009-0003-6128-2347}\\
Luis Colina \orcid{0000-0002-9090-4227} \\
G. Östlin \orcid{0000-0002-3005-1349}\\
Pablo G. P\'erez-Gonz\'alez \orcid{0000-0003-4528-5639}\\
Luca Costantin \orcid{0000-0001-6820-0015}\\
Edoardo Iani \orcid{0000-0001-8386-3546}\\
Pierluigi Rinaldi \orcid{0000-0002-5104-8245}\\
Javier Álvarez-M\'arquez \orcid{0000-0002-7093-1877}\\
A. Bik \orcid{0000-0001-8068-0891}\\
Sarah E. I. Bosman \orcid{0000-0001-8582-7012}\\
Alejandro Crespo Gomez \orcid{0000-0003-2119-277X}\\
Andreas Eckart \orcid{0000-0001-6049-3132}\\
Macarena Garcia Marin \orcid{0000-0003-4801-0489}\\
Thomas R. Greve \orcid{0000-0002-2554-1837}\\
Jens Hjorth \orcid{0000-0002-4571-2306}\\
A. Labiano \orcid{0000-0002-0690-8824}\\
Danial Langeroodi \orcid{0000-0001-5710-8395}\\
J. Melinder \orcid{0000-0003-0470-8754}\\
Florian Peißker \orcid{orcid:0000-0002-9850-2708}\\
Fabian Walter \orcid{0000-0003-4793-7880}\\
M. G\"udel \orcid{0000-0001-9818-0588}\\
Thomas Henning \orcid{0000-0002-1493-300X}\\
Thomas P. Ray \orcid{0000-0002-2110-1068}\\

\bibliography{master}{}
\bibliographystyle{mnras}


\newpage
\appendix

\begin{table*}
\section{Rest-Frame 2~\um\ Morphology}\label{App:RF Morphology}
\renewcommand{\arraystretch}{1.5}
\centering
\caption{Rest-Frame 2~\um\ {\sc{galfit}} and {\sc{statmorph}} outputs, ordered by X-ray ID from \citet{Luo2017}.}
\begin{tabular}{cccccccccc}
\hline
\hline
X-ray ID & $\log_{10}$(L$_{\rm PSF}$) & $\log_{10}$(L$_{\rm Sersic}$) & PSF Lum.  &S\'ersic Index & S\'ersic Size & Gini & M$_{20}$ & Extended & Point Source\\
 & [erg\,s$^{-1}$] & [erg\,s$^{-1}$] & Fraction & ($n$) & ($R_{\rm e})$ [kpc] & &  \\
\hline
667 & 43.26 & 43.13 & 0.57 & 0.69 & 2.85 & 0.51 & $-$1.81 & likely & yes: psf \\
668 & 43.20 & 43.87 & 0.18 & 0.76 & 2.42 & 0.48 & $-$1.55 & yes &  \\
689$^{\rm A}$ & 42.68 & 43.45 & 0.15 & 0.30 & 4.18 & 0.52 & $-$1.75 & yes &  \\
690 & 43.03 & 42.90 & 0.57 & 3.31 & 1.46 & 0.61 & $-$2.09 & yes &  \\
694 & 43.62 & 43.04 & 0.79 & 0.30 & 2.18 & 0.51 & $-$1.72 & likely & yes \\
695 & 36.99 & 40.93 & <0.01 & 0.50 & 2.02 & 0.40 & $-$1.39 & likely &  \\
698 & 43.56 & 44.27 & 0.16 & 2.25 & 4.32 & 0.58 & $-$2.06 & yes &  \\
700 & 43.05 & 43.81 & 0.15 & 2.69 & 3.00 & 0.57 & $-$2.03 & yes & yes: psf \\
703$^{\rm A}$ & 43.43 & 44.10 & 0.18 & 0.37 & 4.51 & 0.46 & $-$1.75 & yes & likely \\
709$^{\rm A}$ & 42.51 & 43.31 & 0.14 & 0.30 & 6.09 & 0.44 & $-$1.04 & likely & likely \\
715$^{\rm A}$ & 44.66 & 44.96 & 0.33 & 2.01 & 0.44 & 0.39 & $-$1.55 &  & yes: psf \\
718$^{\rm A}$ & 43.55 & 44.27 & 0.16 & 0.56 & 1.76 & 0.47 & $-$1.53 & likely &  \\
724$^{\rm A}$ & 42.65 & 42.89 & 0.37 & 2.61 & 0.60 & 0.50 & $-$1.74 &  & yes: psf \\
727 & 42.54 & 43.04 & 0.24 & 5.07 & 2.22 & 0.58 & $-$1.95 & likely & yes: psf \\
735 & 44.43 & 43.76 & 0.83 & 0.57 & 5.37 & 0.56 & $-$1.82 &  & yes: psf \\
745 & 43.57 & 43.84 & 0.35 & 1.45 & 3.18 & 0.57 & $-$1.98 & likely & yes: psf \\
748$^{\rm A}$ & 43.91 & 44.36 & 0.26 & 0.96 & 3.70 & 0.52 & $-$1.90 & yes & yes: psf \\
749$^{\rm A}$ & 42.97 & 43.77 & 0.14 & 0.59 & 4.94 & 0.47 & $-$1.88 & yes & yes \\
751$^{\rm A}$ & 43.70 & 44.42 & 0.16 & 0.87 & 2.16 & 0.56 & $-$1.60 & yes &  \\
754 & 42.22 & 42.33 & 0.44 & 0.76 & 1.57 & 0.51 & $-$1.85 & yes & likely \\
756 & 43.56 & 43.71 & 0.41 & 2.74 & 1.30 & 0.49 & $-$1.77 & yes & likely \\
758 & 42.19 & 43.27 & 0.08 & 0.77 & 2.06 & 0.39 & $-$0.81 & yes &  \\
767 & 42.49 & 43.35 & 0.12 & 3.78 & 2.14 & 0.56 & $-$1.92 & yes & likely \\
788 & 43.73 & 43.67 & 0.54 & 1.16 & 0.79 & 0.41 & $-$1.56 & likely & yes \\
797$^{\rm A}$ & 43.87 & 43.96 & 0.45 & 1.48 & 2.73 & 0.58 & $-$1.92 & yes & yes \\
799 & 42.45 & 42.98 & 0.23 & 0.80 & 2.26 & 0.50 & $-$1.78 & yes &  \\
801 & 43.51 & 43.56 & 0.47 & 0.57 & 3.35 & 0.52 & $-$1.78 & yes & likely \\
805$^{\rm A}$ & 44.38 & 44.44 & 0.46 & 1.66 & 0.68 & 0.47 & $-$1.47 &  & yes: psf \\
809 & 43.11 & 43.09 & 0.51 & 0.86 & 1.15 & 0.51 & $-$1.72 & likely & likely \\
840 & 43.40 & 42.78 & 0.81 & 0.39 & 6.33 & 0.55 & $-$1.76 & likely & yes: psf \\
S-36 & 41.36 & 42.23 & 0.12 & 0.51 & 1.65 & 0.45 & $-$1.70 & yes & likely \\
\hline
\hline
\end{tabular}
\tablefoot{Column 1 lists the IDs adopted from \citet{Luo2017}, where an `A' indicates the source is detected in the ASPECS ALMA survey of the HUDF \citep{Boogaard2024}. Columns 2 and 3 indicate the rest-frame 2~\um\ luminosities for the PSF and S\'ersic model, converted to units of erg\,s$^{-1}$ using the rest-frame frequency, whilst column 4 details the PSF to total luminosity fraction of the source from the {\sc{galfit}} modelling. The S\'ersic index, effective radius, Gini and M20 parameters are given in columns 5, 6, 7 and 8. In columns 9 and 10 we list the visual classification of extended and point-source components, for ease of comparison with the quantitative results (see Appendix \ref{sect_visClass} for details).}
\end{table*}

\FloatBarrier


\FloatBarrier

\begin{table*}
\section{Visual classification: details} \label{sect_visClass}
\caption{Visual classification for each object in our sample, ordered by X-ray ID from \citet{Luo2017}.
}             
\label{Table:VisClass}
\centering
\renewcommand{\arraystretch}{1.5}
\begin{tabular}{l l l l l l l l p{4cm} }     
\hline
\hline
ID & Class & Class\_ref & \multicolumn{5}{c}{Morphological categories identified in each object} &  \\
   &       &            & Disk & Spheroid & Irregular & Extended & Point\_Source & Notes \\
   &       &            &      &          &           &          & [c]           &       \\
\hline
667	&	AGN	&	Luo17	&		&	likely	&		&	likely	&	yes: psf	&		\\
668	&	AGN	&	Lyu22	&	likely	&	yes	&		&	yes	&		&	00923 [a]	\\
689	&	AGN	&	Luo17	&	yes	&		&	likely	&	yes	&		&	interacting?	\\
690	&	Galaxy	&	Luo17	&	likely	&	yes	&	likely	&	yes	&		&	interacting?	\\
694	&	Galaxy	&	Luo17	&		&	likely	&		&	likely	&	yes	&	interacting?	\\
695	&	AGN	&	Lyu22	&	likely	&		&	likely	&	likely	&		&	interacting?	\\
698	&	AGN	&	Lyu22	&		&	yes	&		&	yes	&		&		\\
700	&	AGN	&	Lyu22	&		&	yes	&		&	yes	&	yes: psf	&		\\
703	&	AGN	&	Lyu22	&	yes	&	yes	&		&	yes	&	likely	&	01520 [a]	\\
709	&	Galaxy	&	Luo17	&	likely	&	likely	&	likely	&	likely	&	likely	&	interacting?;  East IR source is X-ray source	\\
715	&	AGN	&	Luo17	&		&		&		&		&	yes: psf	&		\\
718	&	AGN	&	Luo17	&		&	likely	&	likely	&	likely	&		&	interacting?	\\
724	&	Galaxy	&	Luo17	&		&		&		&		&	yes: psf	&		\\
727	&	AGN	&	Lyu22	&	likely	&	likely	&		&	likely	&	yes: psf	&		\\
735	&	AGN	&	Luo17	&		&		&		&		&	yes: psf	&		\\
745	&	AGN	&	Luo17	&		&	likely	&		&	likely	&	yes: psf	&		\\
748	&	AGN	&	Luo17	&	yes	&	likely	&		&	yes	&	yes: psf	&		\\
749	&	Galaxy	&	Luo17	&	yes	&	likely	&		&	yes	&	yes	&		\\
751	&	AGN	&	Lyu24	&		&		&	yes	&	yes	&		&	interacting?; 00272 [a]; complex morphology	\\

754	&	AGN	&	Luo17	&	yes	&	yes	&		&	yes	&	likely	&	interacting?	\\
756	&	AGN	&	Lyu22	&		&	yes	&		&	yes	&	likely	&		\\
758	&	AGN	&	Luo17	&		&		&	yes	&	yes	&		&	interacting?	\\
767	&	AGN	&	Lyu22	&	likely	&	yes	&		&	yes	&	likely	&		\\
788	&	AGN	&	Luo17	&		&	likely	&	likely	&	likely	&	yes	&	interacting?; [b]	\\
797	&	AGN	&	Luo17	&		&	yes	&		&	yes	&	yes	&	interacting?	\\
799	&	Galaxy	&	Luo17	&	yes	&	yes	&		&	yes	&		&		\\
801	&	AGN	&	Lyu22	&		&	yes	&	likely	&	yes	&	likely	&	interacting?	\\
805	&	AGN	&	Luo17	&		&		&		&		&	yes: psf	&	interacting?	\\
809	&	AGN	&	Luo17	&		&	likely	&	likely	&	likely	&	likely	&	interacting?	\\
840	&	AGN	&	Luo17	&		&	likely	&		&	likely	&	yes: psf	&		\\
S-36	&	Galaxy	&	Luo17	&	yes	&	yes	&		&	yes	&	likely	&	00862 [a]	\\
\hline  
\hline
\end{tabular}
\tablefoot{ \\
a. NGDEEP\_nnnnn (Shen+24 H-alpha map)  \\
b. also in $z>3$ massive galaxies sample (CANDELS 14587; Costantin+24) \\
c. yes: psf -- indicates PSF characteristic diffraction pattern is visible \\
}
\end{table*}

\begin{figure*}
  \section{Colour Maps}\label{App:CMaps}
    \centering
    \includegraphics[width=\linewidth,trim={0cm 3cm 0cm 0cm}]{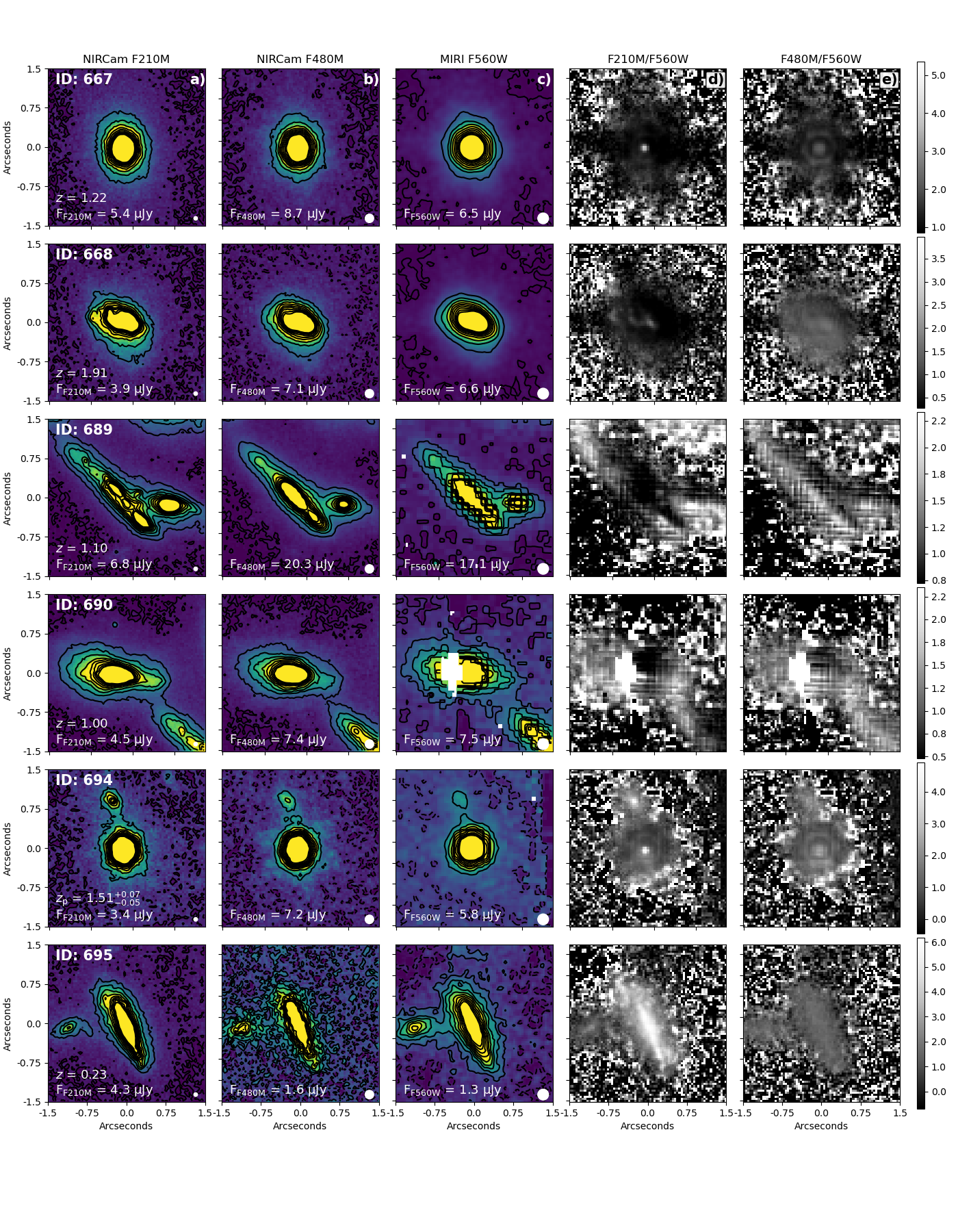}
    \caption{Multi-wavelength imaging and colour maps for all 31 X-ray sources in our sample. For each galaxy, we show, and contour, a 3-arcsecond cutout in the NIRCam F210M band (a), NIRCam F480M band (b) and the MIRI F560W band (c). For each cutout, we label the galaxy's ID as given by \citet{Luo2017}, measured flux in each band and spectroscopic redshift. We also indicate the FWHM of the PSF in each band by the white circle in the lower-right corner. The F210M/F560W (d) and F480M/F560W (e) colour maps are shown for each source, highlighting the presence of any AGN (point source) component at longer (near-infrared) wavelengths.}
\end{figure*}

\begin{figure*}
  
    \centering
    \includegraphics[width=\linewidth,trim={0cm 3cm 0cm 0cm}]{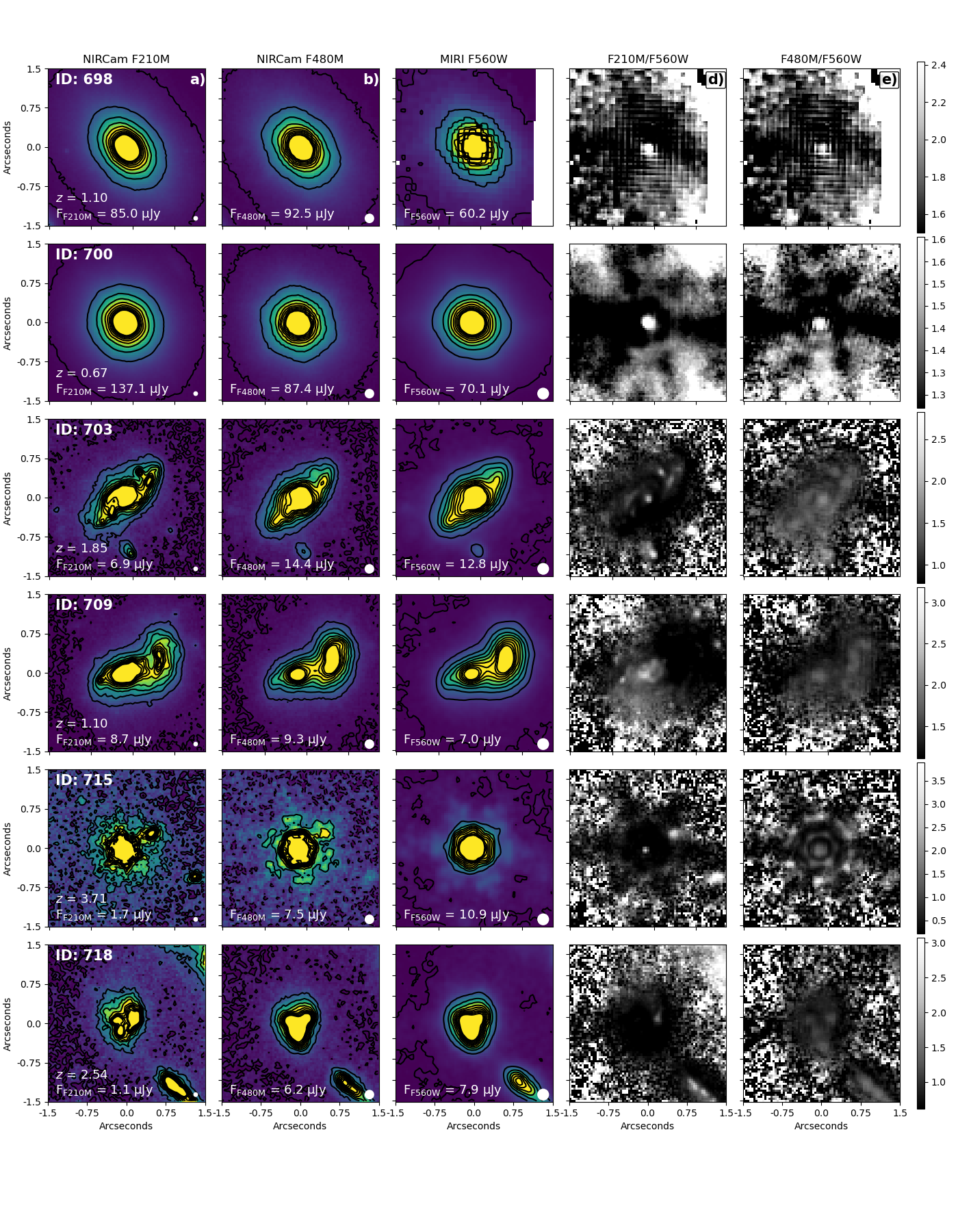}
    \caption{Continued....}
\end{figure*}

\begin{figure*}
   
    \centering
    \includegraphics[width=\linewidth,trim={0cm 3cm 0cm 0cm}]{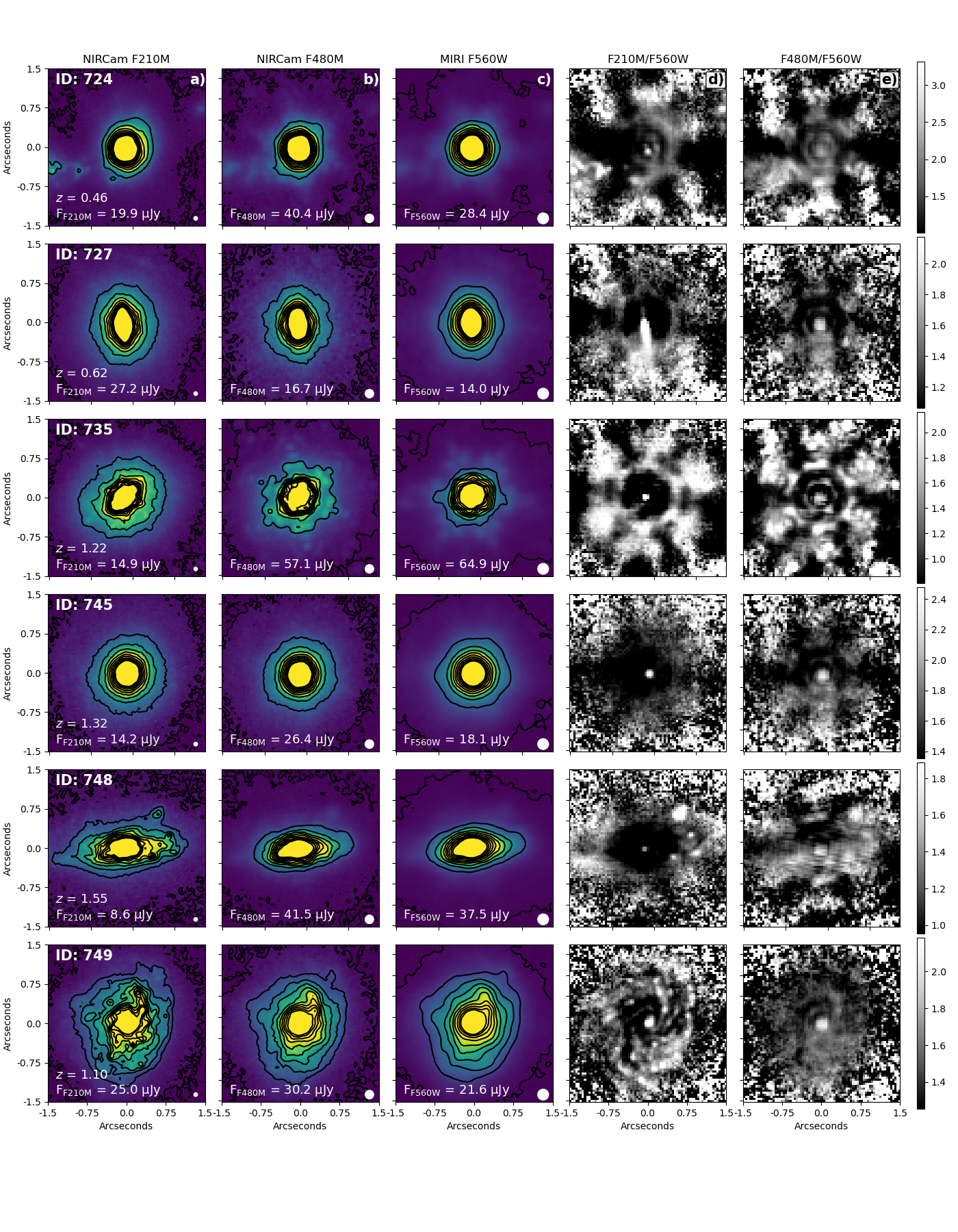}
    \caption{Continued....}
\end{figure*}

\begin{figure*}

    \centering
    \includegraphics[width=\linewidth,trim={0cm 3cm 0cm 0cm}]{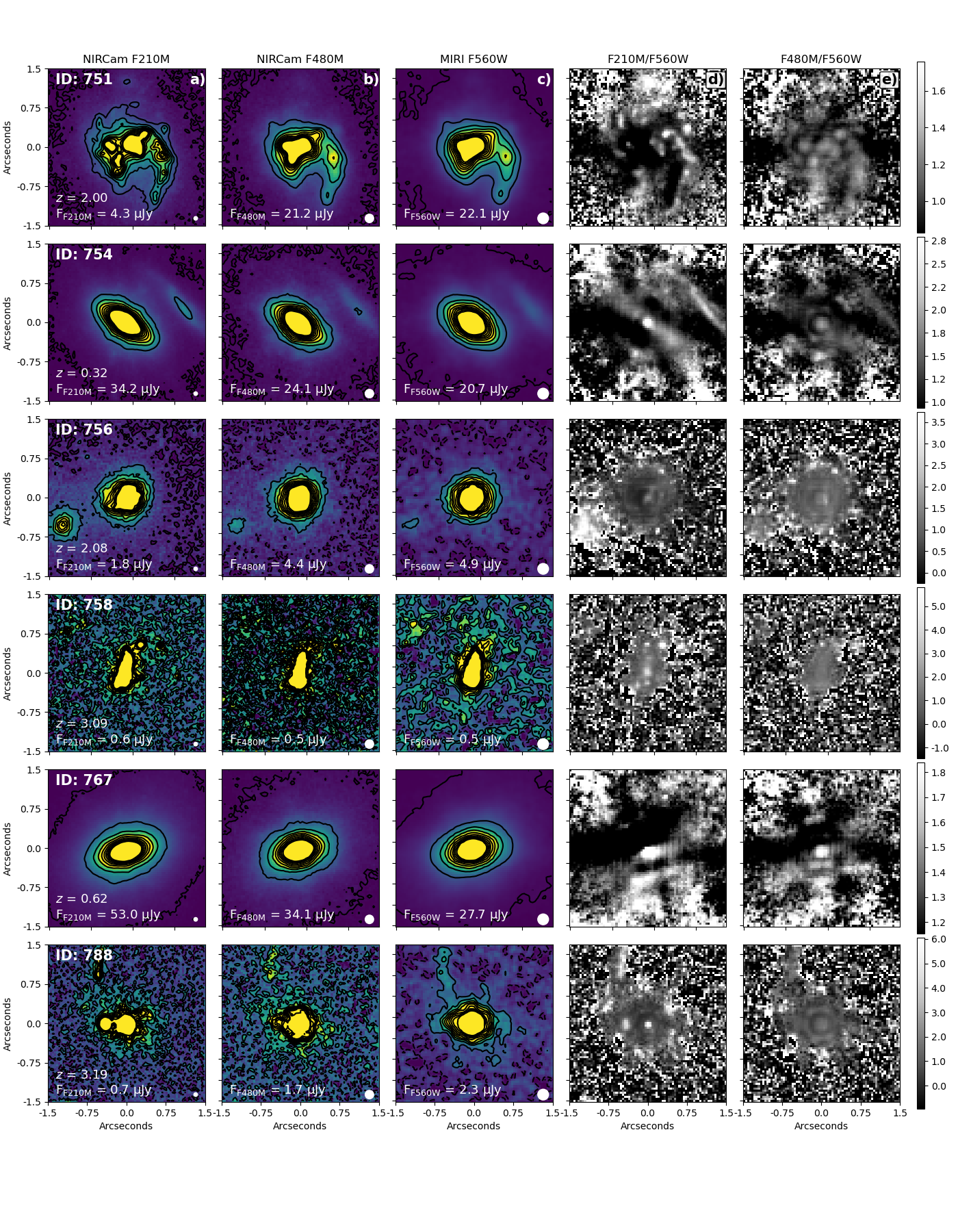}
    \caption{Continued....}
\end{figure*}

\begin{figure*}
 
    \centering
    \includegraphics[width=\linewidth,trim={0cm 3cm 0cm 0cm}]{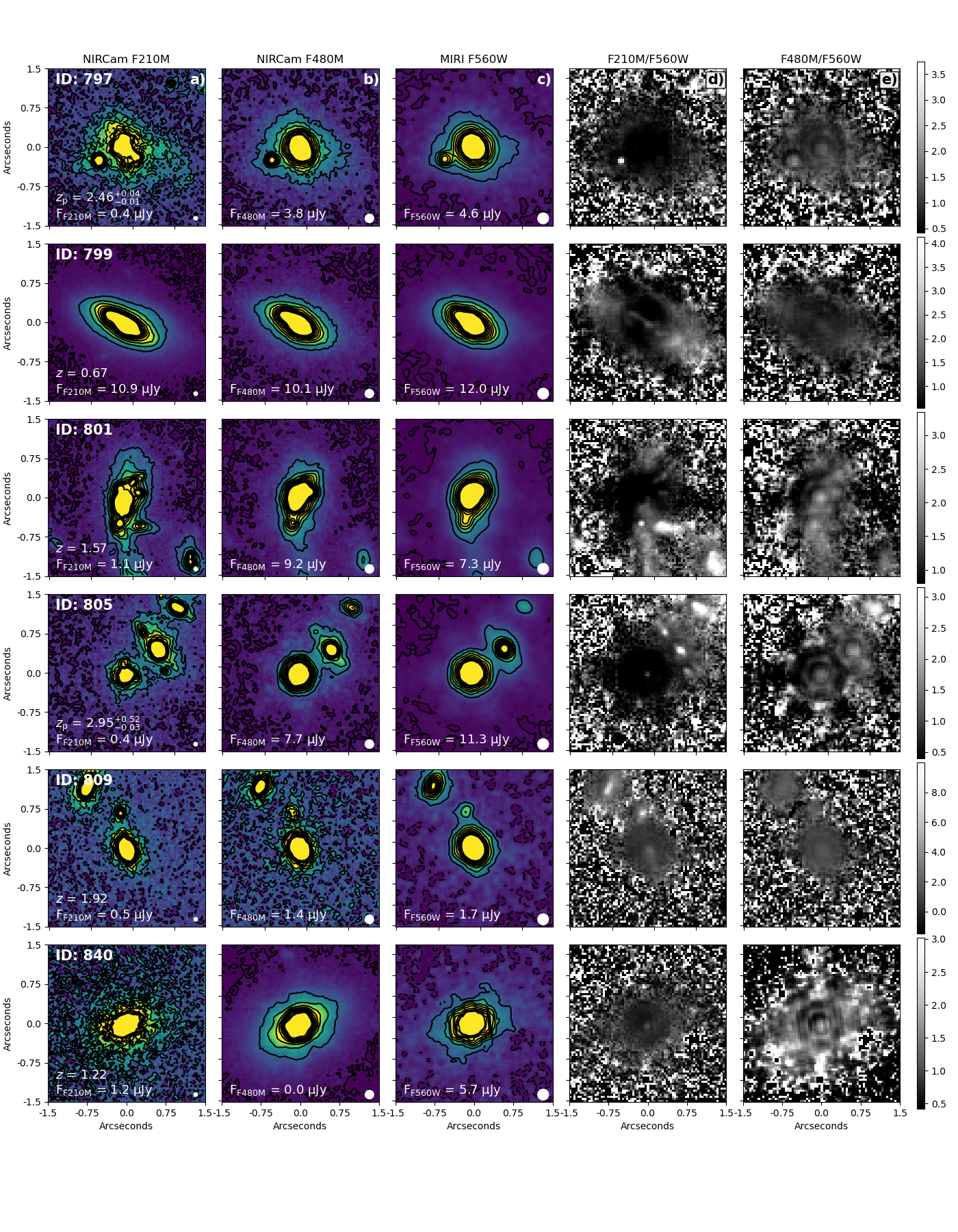}
    \caption{Continued....}
\end{figure*}

\begin{figure*}
   
    \centering
    \includegraphics[width=\linewidth,trim={0cm 38cm 0cm 0cm},clip]{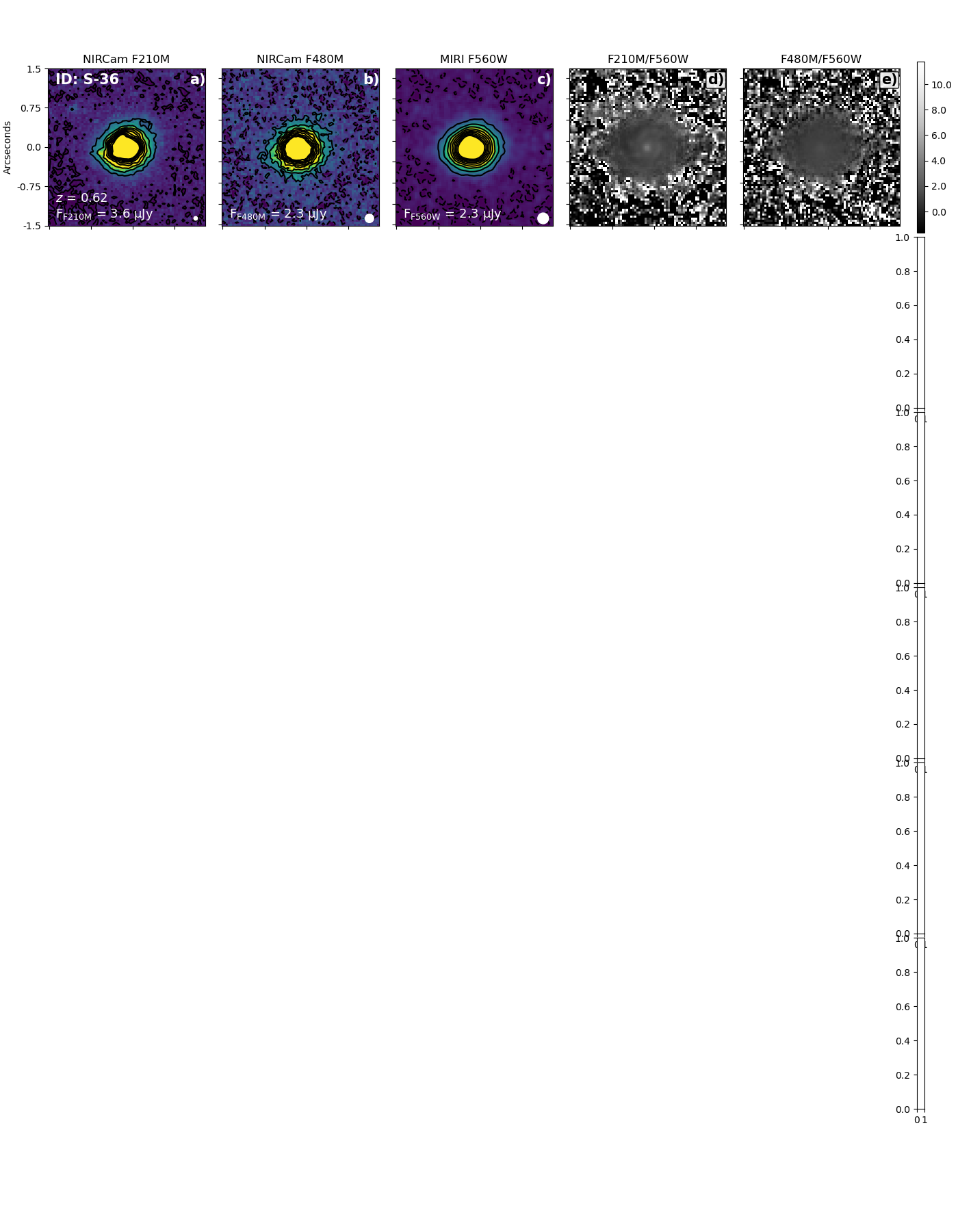}
    \caption{Continued....}
\end{figure*}

\begin{figure*}
  \section{Morphology Statistics}\label{App:GALFIT}
    \centering
    \includegraphics[width=\linewidth]{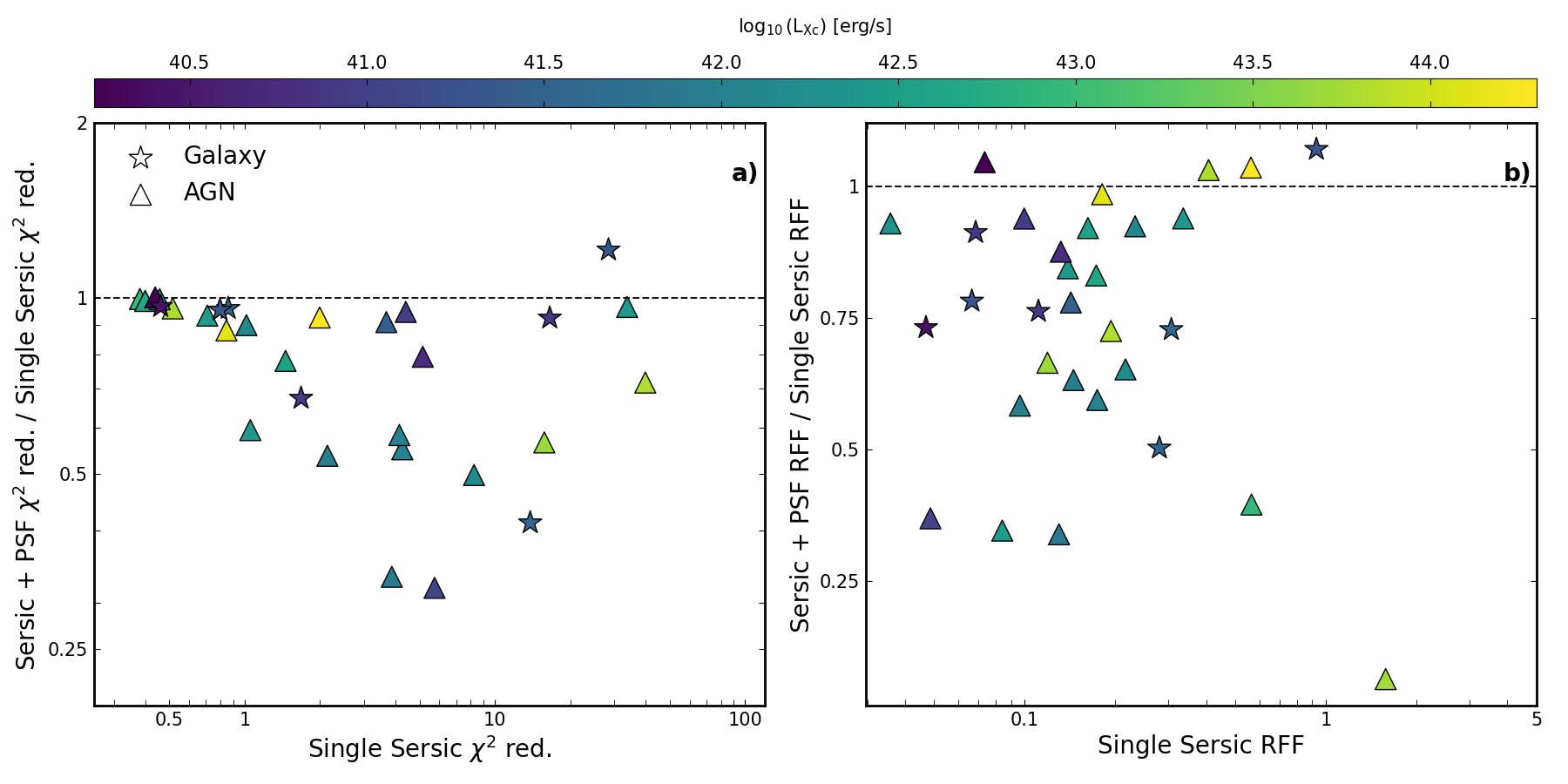}
    \caption{The {\sc{galfit}} modelling statistics for the 31 X-ray sources at rest-frame 2~\um. We compare the single S\'ersic fit to a S\'ersic plus PSF model, analysing the difference in reduced chi-squared {\it{(a)}}, and RFF {\it{(b)}}. For the majority of sources, we identify improved reduced chi-squared and RFF values between the two models.}
    \label{Fig:Morph_Fit_Stats}
\end{figure*}


\end{document}